\documentclass[aip,pof,reprint,floatfix]{revtex4-1}
\usepackage[dvipsnames,rgb,dvips]{xcolor}
\usepackage{graphicx}
\usepackage{dcolumn}
\usepackage{float}
\usepackage{overpic}
\usepackage{amsmath,amssymb}
\usepackage{bm}
\usepackage[]{units}
\usepackage{upgreek}
\newcommand{\obs}[1]{{#1}}
\newcommand{\ve}[1]{\ensuremath{\mbox{\boldmath$#1$}}}
\newcommand{\ma}[1]{\ensuremath{\mathbb{#1}}}

\begin{document}

\title{Tumbling of asymmetric microrods in a microchannel flow}
\author{J. Einarsson}
\email[]{These authors  contributed  equally to this work.}
\author{B. M. Mihiretie}
\email[]{These authors  contributed  equally to this work.}
\affiliation{Department of Physics, University of Gothenburg, 41296 Gothenburg, Sweden.}
\author{A. Laas}
\affiliation{Department of Physics, University of Gothenburg, 41296 Gothenburg, Sweden.}
\author{S. Ankardal}
\affiliation{Department of Physics, University of Gothenburg, 41296 Gothenburg, Sweden.}
\author{J. R. Angilella}
\affiliation{LUSAC, Universit\'e de Caen, Cherbourg, France.}
\author{D. Hanstorp}
\affiliation{Department of Physics, University of Gothenburg, 41296 Gothenburg, Sweden.}
\author{B. Mehlig}
\affiliation{Department of Physics, University of Gothenburg, 41296 Gothenburg, Sweden.}

\date{\today}

\begin{abstract} 
We describe results of measurements of the orientational motion of glass microrods in a microchannel flow,  following the orientational motion of particles with different  shapes. We determine how the orientational dynamics depends  on the shape of the particle and on its initial orientation. We find that the dynamics depends so sensitively on the degree to which axisymmetry is broken that it is difficult to find particles that are sufficiently axisymmetric so that they exhibit periodic tumbling (\lq Jeffery orbits\rq{}). \obs{The results of our measurements confirm earlier theoretical analysis predicting sensitive dependence on particle shape and its initial orientation.  Our results  illustrate the different types of orientational dynamics for asymmetric particles predicted by theory.}
\end{abstract}
\maketitle

\section{Introduction}
\label{sec:intro}

We study experimentally the orientational dynamics of neutrally buoyant non-axi\-sym\-me\-tric particles suspended in a viscous shear flow. 
\obs{The rotation of axisymmetric particles in a shear flow has  been studied in several experiments (we give a brief account below).
The case of non-axi\-sym\-me\-tric particles, by contrast, has received little attention experimentally. This is surprising because it is known in theory\cite{Hinch1979,Yarin1997} that the orientational dynamics can be very sensitive to small deviations from the axisymmetric limit. How sensitively the dynamics is affected by slight breaking of axisymmetry depends upon the orientational trajectory which in turn is determined by the initial orientation.}

\obs{In order to verify this theoretical prediction experimentally it is necessary to use particles with well-defined shapes, to make sure that inertial effects and thermal noise are negligible, and to compare different orientational trajectories of the same particle.}

\obs{In this paper we describe experimental observations of the orientational motion of micron-sized glass particles suspended in a pressure-driven micro-channel flow.  The particles have different shapes: axisymmetric, slightly non-axisymmetric, and substantially non-axisymmetric (the latter are 
strongly triaxial particles made for the purpose of this experiment by joining two micron-sized glass rods). We verify that inertial and thermal torques have negligble effects by showing that the orientational dynamics is invariant under reversal of the pressure. 
An optical trap allows us to manipulate the same particle into different initial orientations, to study different orientational
trajectories of the same particle. The results of our measurements confirm the predictions of Refs.~\citenum{Hinch1979,Yarin1997}, show
the sensitive dependence of the orientational dynamics upon particle shape and initial orientation, and qualitatively illustrate the different types of orientational dynamics computed in Refs.~\citenum{Hinch1979,Yarin1997}. 
We have not attempted to quantititatively compare individual experimental
particle trajectories with theory,
to compare individual trajectories 
requires to compute the resistance tensors for the actual shape of the particles used in the experiments
(the theory of Refs. \citenum{Hinch1979,Yarin1997} is formulated for ellipsoidal particles).  This is beyond the scope of the present paper. Besides the resulting trajectories are of little general interest, they depend very
sensitively on the particular  shape of the particle in question.
}

\obs{The remainder of this Introduction briefly introduces the relevant theory as well as the wider context of this work.} 
The orientational dynamics of a particle suspended in a viscous flow is determined by resistance tensors that relate the local flow velocity and its gradients to the torque acting on the particle\cite{Happel1965,Kim1991}. 
\obs{A given particle shape corresponds to a set of resistance tensors. Their elements are computed by solving Stokes' equations in appropriate geometries\cite{Happel1965,Kim1991}. The case of an ellipsoid in a viscous shear flow was first solved by Jeffery\cite{Jeffery1922}.}
\obs{In this limit} the equation of motion is given by the condition that the \obs{hydrodynamic} torque vanishes at every instant. For particles that possess three orthogonal mirror planes\cite{Bretherton1962}, particle shape enters the orientational equation of motion through two parameters, $\Lambda$ and $K$. For an ellipsoid with half-axes $a$, $b$ and $c$, for example, $\Lambda=(\lambda^2\!-\!1)/(\lambda^2\!+\!1)$ and $K=(\kappa^2\!-\!1)/(\kappa^2\!+\!1)$ with aspect ratios $\lambda=a/c$ and $\kappa=b/c$.

The case $K=0$ corresponds to an axisymmetric particle. For axisymmetric particles the orientational dynamics is exactly solvable\cite{Jeffery1922}. 
When $|\Lambda|<1$ (so that $0 < \lambda < \infty$) 
there are infinitely many degenerate periodic orbits, the so-called \lq Jeffery orbits\rq. This is a consequence of the fact that the dynamical system has a conserved quantity: the \lq orbit constant\rq. The value of the orbit constant is determined by the initial orientation of the particle. The orientational motion of an axisymmetric particle in a simple shear is sometimes referred to as \lq tumbling\rq{}, the particle spends a long time aligned with the flow direction, and it periodically changes orientation by $180$ degrees. Different Jeffery orbits differ in the functional form of these periodic \lq flips\rq{}.
The  degeneracy for $K=0$ is particular to the simple shear flow, and it means that small perturbations can have a large effect. 

Inertial forces, for example, are neglected in Jeffery's theory. These forces induce \lq orbit drift\rq{} into a final stable orbit.  This was already suggested by Jeffery\cite{Jeffery1922}, and was discussed in many experimental papers starting with Taylor\cite{Taylor1923}. See also Ref.~\citenum{Karnis1966}. 
\obs{The corresponding theory is discussed by \citet{subramanian2005} and by Einarsson {\em et al.}\cite{einarsson2015a,einarsson2015b,candelier2015}.}

Small particles may be affected by thermal noise so that the orbit constant performs a random walk giving rise to a statistical distribution of orientations. This mechanism  forms the theoretical basis for understanding the rheology of dilute suspensions\cite{Brenner1974,Hinch1972}.

A third possibility, the topic of this work, is that the particle is not perfectly axisymmetric. This leads to a more complicated orientational equation of motion, also derived by Jeffery\cite{Jeffery1922}. Some numerical examples of its solutions were reported by \citet{Gierszewski1978} who found that the motion of a non-axisymmetric particle in a simple shear flow is qualitatively different from that of an axisymmetric particle. \citet{Hinch1979} analysed the structure of the solutions to the equation of motion.
They found that for short times a nearly axisymmetric ellipsoidal particle approximately follows a Jeffery orbit, but on longer time scales  the \lq orbit constant{\rq} does not remain
constant. It oscillates, giving rise to \lq doubly periodic\rq{} tumbling: time series of the components of the unit vector aligned with the major axis of the particle  show two distinct periods. Subsequently Yarin {\em et al.}\cite{Yarin1997} inferred from numerical experiments and analytical calculations
that ellipsoidal particles may tumble periodically, quasi-periodically, or in a chaotic fashion -- depending on the particle shape and on its initial orientation. The term \lq quasi-periodic\rq{} refers to doubly periodic motion with incommensurable periods. Our experimental results support these theoretical predictions: our analysis demonstrates that the tumbling may indeed be periodic, doubly periodic, or possibly chaotic, depending on particle shape and initial orientation.  Our results are in good qualitative agreement with theoretical predictions.

\obs{This work considers the orientational motion of small neutrally buoyant particles in a time-independent viscous shear flow.}  This is a special but important case. It is of theoretical interest because of its degeneracy and sensitivity to small perturbations, and it is of practical interest because it fundamentally relates to theories and experiments concerning the rheology of dilute suspensions. Theories are commonly formulated in terms of Jeffery's equation\cite{Brenner1974,Petrie1999}. 
Recently there has been a surge of interest in describing the tumbling of non-spherical particles in turbulent\cite{Par12,Pum11,Gus14,Ni14,Che13,Byron2015} and other complex flows\cite{Wil09,Wil10a,Wil11}.
Since it is difficult to solve the coupled particle-flow problem most theoretical and numerical studies rely 
on Jeffery's equation as an approximation to  the orientational dynamics. 
Some exceptions are described in Refs.~\citenum{Marchioli2010,Einarsson2013a,Challabotla2015}.

This article is organised as follows. In Section~\ref{sec:background} we enumerate previous experimental efforts to validate Jeffery's equations. In Section~\ref{sec:methods} we describe the experimental setup. Section~\ref{sec:results} contains our experimental results. These results are discussed in Section~\ref{sec:discussion}, and we conclude in Section~\ref{sec:conclusions}.

\section{Previous experimental work}\label{sec:background}
In this section we give a brief account of earlier experiments observing the orientational dynamics of single particles in shear flows.

\citet{Taylor1923} immersed millimeter-sized aluminum spheroids in sodium silicate between two concentric rotating cylinders approximately \unit[10]{mm} apart. In his brief report he asserts that the tumbling of the spheroids is in qualitative agreement with Jeffery's predictions, but that the orientational dynamics drifts after many (order of $100$) particle rotations.

In a related study Eirich {\em et al.}\cite{Eirich1936} observed the orientations of glass rods and silk fibres in a Taylor-Couette device.    
The particles had diameters between $10$ and \unit[50]{$\upmu$m} and aspect ratios between $5$ and $100$. No quantitative data on the orientational dynamics was measured, but they observed that the particles tend to align with the flow direction or 
with the vorticity direction.

\citet{Binder1939} studied fibers of many different aspect ratios suspended in glycerine. 
He employed a similar device with two concentric cylinders and found, as Taylor, that the orientational dynamics
slowly drifts. 

Mason and co-workers have studied the orientational dynamics of small particles in shear flows during two decades\cite{Trevelyan1951,Mason1956,Bartok1957,Goldsmith1961,Goldsmith1962,Anczurowski1968}.  Initially \citet{Trevelyan1951} used a setup of two concentric cylinders rotating in opposite directions, making it possible to study a single particle over an extended period of time. The gap between the cylinders was approximately \unit[10]{mm}, the suspending liquid was white corn syrup, and the particles were \unit[9.5]{$\upmu$m} diameter glass fibers cut to different lengths. By observing the particle orientation in a plane orthogonal to vorticity, \citet{Trevelyan1951} 
found fairly good quantitative agreement of their experimental results with Jeffery's theory for one particle rotation (Fig.~7 in Ref.~\citenum{Trevelyan1951}). However, for longer time series (up to 30 revolutions) their results were inconclusive: sometimes orbit drift was observed, sometimes not, and sometimes the change of orbit appeared seemingly erratic. 
In order to compare quantitatively with Jeffery's equations, valid for spheroidal particles, Trevelyan and Mason  fitted the value of $\Lambda$ to measurements of the tumbling period, yielding in their words an \lq effective aspect ratio\rq{} for cylinders. 
\citet{Bretherton1962} later showed that this procedure is consistent.
\citet{Mason1956} extended the experiment to hundreds of particle rotations (Fig.~2 in Ref.~\citenum{Mason1956}), but the observed orbit drift was apparently erratic. 
Mason and Manley mention convective currents as a possible cause for the observed drift, but no conclusions could be drawn concerning the single-particle dynamics.
\citet{Bartok1957} used a similar device consisting of concentric cylinders, with a camera-equipped microscope observing along the vorticity axis, allowing
to very precisely measure the tumbling behavior of high aspect ratio ($\lambda\approx45)$ acrylic (\lq Orlon\rq) fibres. The experimental results
were found to agree quantitatively with Jeffery's predictions for one particle rotation (Fig.~5 in Ref.~\citenum{Bartok1957}). However, no data on the orbit drift, if any, was presented.
\citet{Goldsmith1961} performed the first quantitative measurements of the rotations of disks. They used the same coaxial-cylinder setup described above, 
with silicone oil for the suspending liquid. The disks were fabricated by heating and compressing 
polystyrene spheres. The disk diameters were \unit[400-850]{$\upmu$m}, 
and their aspect ratios ranged from $\lambda=1/20$ to $\lambda = 1/4$. Goldsmith and Mason showed that Jeffery's theory quantitatively predicts the orientational motion of 
axisymmetric disks, and that the orbit remained constant over $120$ particle rotations (Table~II in Ref.~\citenum{Goldsmith1961}).
In a sequel \citet{Goldsmith1962} described measurements of the motion of particles suspended in a liquid flowing through a circular tube. The tube diameters were \unit[2-8]{mm}, the flow was pressure-driven by a syringe pump, and observations were recorded through a microscope traveling along the tube. Measurements were made on an assortment of particles 
of sizes $\approx$\unit[0.1]{mm} with aspect ratios ranging from $\lambda=1/20$ (disks) to $\lambda=100$ (fibres). Goldsmith and Mason concluded that the dynamics along 
Jeffery orbits compares well with theory for short times (Fig.~5 in Ref.~\citenum{Goldsmith1962}). However, they found measurable orbit drift after a single particle rotation, for both a rod 
and a disk,
see Fig.~7 in the same paper. The drift was attributed to particle asymmetry.
\citet{Anczurowski1968} fabricated prolate spheroidal particles by polymerising an electrostatically deformed droplet, and showed that Jeffery's theory holds quantitatively for one particle rotation given the true aspect ratio $\lambda$. They used the same concentric-cylinder device described above.

Harris {\em et al.}\cite{Harris1979} performed experimental measurements on non-axi\-sym\-me\-tric particles ($K\neq0$). They used an apparatus with counter-rotating cylinders with a gap of \unit[27]{mm}, which was filled with a glucose solution. The particles were machined from a composite material (\lq Tufnol\rq) into cuboids of $\unit[1.75]{mm}\!\times\!\unit[1.28]{mm}$ cross-section and \unit[2.5-9.5]{mm} in length. They measured           the unknown elements of the resistance tensors by observing simple rotations around each of the principal axes. With these numerical values of the resistance tensors they compared orientational trajectories of the cuboids to numerical solutions of Jeffery's equations and found reasonable quantitative agreement for two particle revolutions (Fig.~9 in their paper).

\citet{Stover1990} investigated the effect of a wall on fibre motion using a pressure-driven flow of corn syrup through a rectangular channel. The fibres had cross-sectional diameters of \unit[50]{$\upmu$m} and lengths \unit[600]{$\upmu$m}, resulting in an aspect ratio of $12$. They found that the orientational dynamics are in quantitative agreement with Jeffery's theory for two particle rotations when the particle is at least one particle length away from the wall.

\citet{Kaya2009} observed \emph{E. coli} bacteria advected in a microfluidic channel of rectangular cross section. 
They found that the orientational motion of the bacteria approximately follows Jeffery orbits.

\citet{Einarsson2013a} described examples of orientational trajectories of polymer microrods in a microchannel. They found that the trajectories of some particles were  periodic, admitting comparison to Jeffery's theory. 
Other trajectories were seen to be aperiodic \obs{(Fig.~8 of Ref.~\citenum{Einarsson2013a}). The authors suggested
that this aperiodic motion may have been caused by lack of axisymmetry
of the particle in question. But it was not possible to draw definite conclusions, for several reasons. First the particles were 
produced by shearing polymer microspheres\cite{alargova2004} which does not produce sufficiently well-controlled shapes. This is a problem because the shape of a given particle in the channel cannot be accurately observed. Second it could not be excluded that thermal torques affected the orientational dynamics. Third, and most importantly, the setup did not allow to record different orientational trajectories of the same particle. This motivated us to perform the experiments summarised in the following, they overcome the problems listed above.}

\section{Methods}
\label{sec:methods}
\begin{figure}[t]
\begin{overpic}[width=8cm,clip]{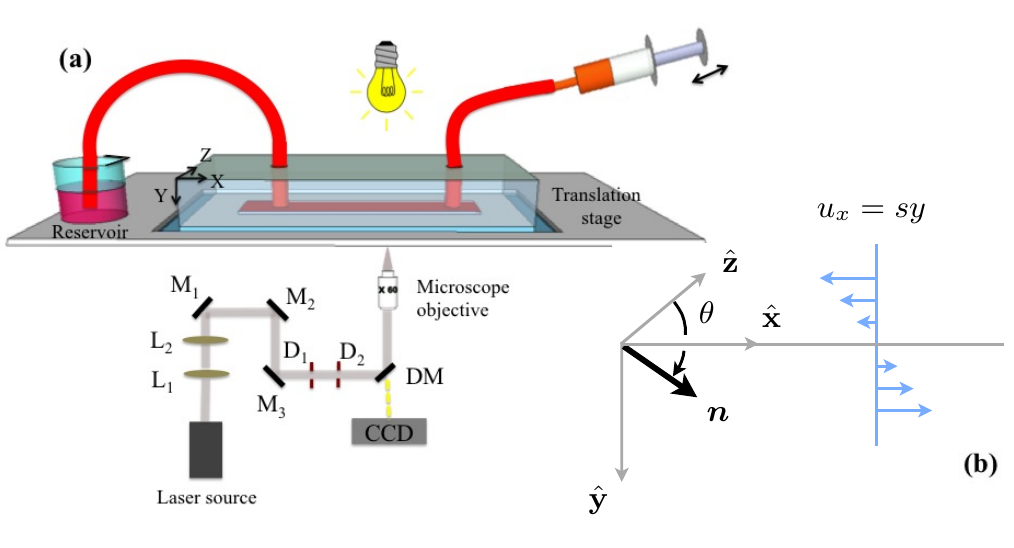}
\end{overpic}
\caption{\label{fig:exp_setup} (a) Schematic picture of the experimental setup, 
elements are not drawn to scale. $L_{1,2}$: lenses in a Keplerian telescope configuration. $M_{1,2,3}$: mirrors. D$_{1,2}$: diaphragms. DM: dichroic mirror 
reflecting the laser beam vertically towards the microfluidic system. The microscopic objective focuses the beam on the sample plane. Illumination is provided from the top.
(b) Coordinate system in the lab frame spanned by orthogonal unit vectors $\hat{\bf x}$, $\hat{\bf y}$, and $\hat{\bf z}$. The $x$-axis (flow direction) is  directed along the channel length, the $y$-axis along the depth of the channel, and the $z$-axis  along the channel width. The unit vector $\ve n$ points
along the major axis of the particle.  The polar angle between $\ve n$ and the ${z}$-axis is denoted by $\theta$. 
The particles are kept roughly equally far away from the side walls of the channel. Since the channel is much wider than deep this means that the $\hat{\bf y}$-direction
is the shear direction, so that the local linearisation of the flow-velocity 
obeys $u_x = sy$ in the frame co-moving with the centre-of-mass
of the particle, where $s$ is the shear rate.}
\end{figure}

Fig.~\ref{fig:exp_setup}(a) shows a schematic drawing of the experimental setup to observe the orientational  motion of small particles advected 
in a microchannel flow. A dilute suspension of microrods in a density-matched fluid is introduced into the microchannel. A syringe pump (a standard Harvard  Apparatus infuse/withdraw model) is used to drive the flow of the particles. The system is placed under an inverted microscope equipped with a motorised translation stage. A CCD camera is used to register the orientation of a given  particle. The initial orientation and position of the particle in the channel is set using an optical trap.     

Fig.~\ref{fig:exp_setup}(b) shows the coordinate system that is used in this paper. The $x$-axis is parallel to the channel length along the flow direction. The $y$-axis is directed along the depth of the channel. The $y$-axis is also the optical axis of the microscope objective. The $z$-axis points along the channel width. The orientation of the particle is determined by the unit vector $\ve n$ along the major axis of the particle.

\begin{figure}[t]
\includegraphics[width=8cm]{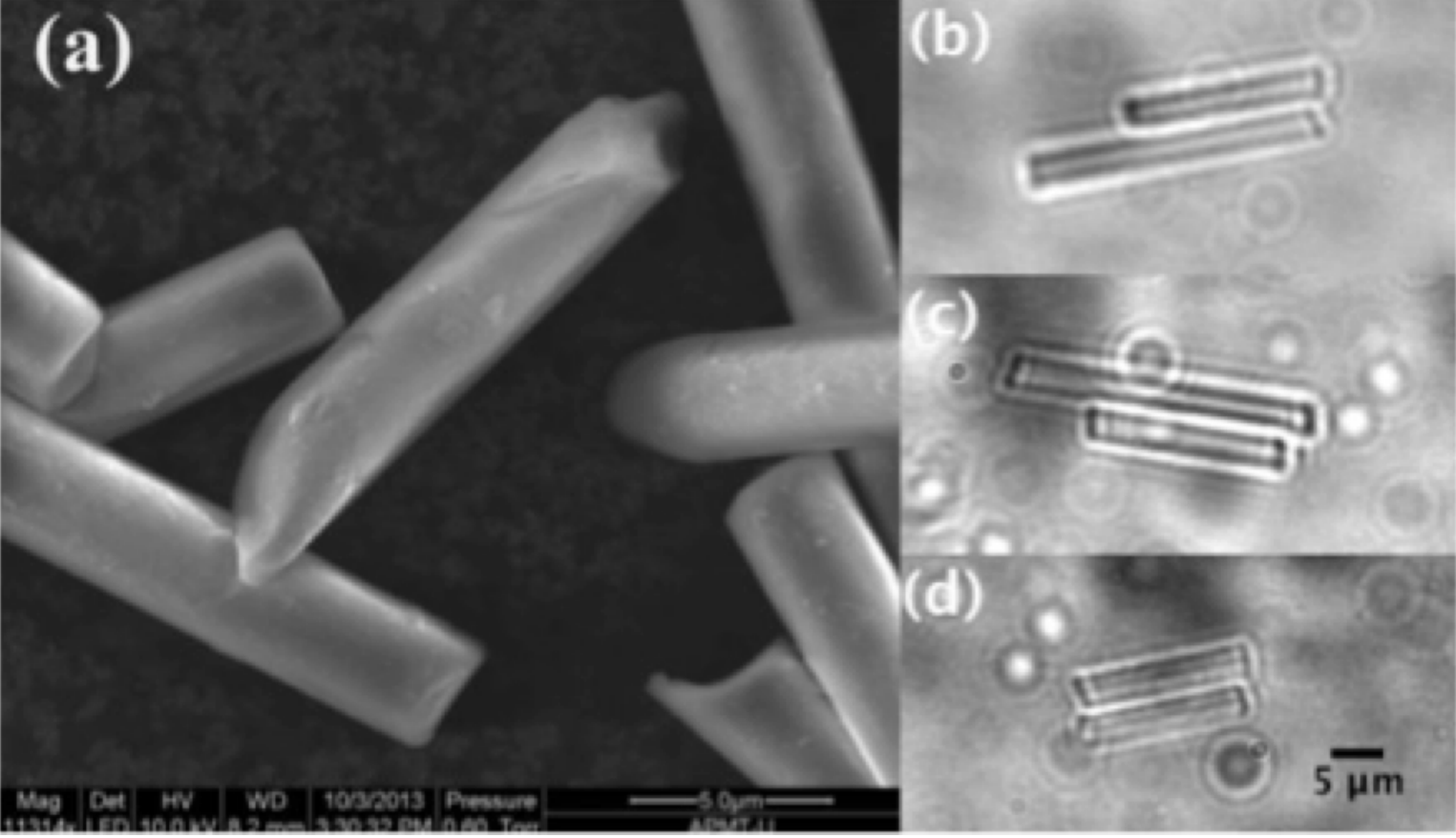}
\caption{\label{fig:particle_image}  (a) Electron-microscope image of the glass particles used in the experiments. Taken by S. Gustafsson, Chalmers.
\obs{(b-d) Optical-microscope images of the double particles used
in the experiments.  The angular dynamics of the particle in panel (b) 
is shown in Fig.~\ref{fig:particle3}, 
panel (c) corresponds to Fig.~S8, and panel (d) to Fig.~S9.
The last two figures are found in the Supplementary Online Material \cite{supp}.}}
\end{figure}

The experiment is performed with cylindrical glass rods with diameters \unit[3]{$\upmu$m} $\pm$ \unit[0.01]{$\upmu$m} (PF-30S, Nippon Electric Glass Co., Ltd). 
The microrods were manufactured as spacers in liquid-crystal devices (PF-30S, Nippon Electric Glass Co., Ltd). This requires precise diameters. 
The lengths of the rods vary between approximately \unit[10]{$\upmu$m} and \unit[30]{$\upmu$m}.
An electron-microscope image of the particles is shown in Fig.~\ref{fig:particle_image}(a). 
This Figure shows that the end surfaces of the cylindrical rods are randomly 
inclined and uneven, indicating that the particles were obtained
by breaking longer glass rods. While this is irrelevant for the intended industrial use as spacers,
it is important for our application. Random inclinations of the end surfaces break axisymmetry: sometimes only very slightly
[lower left particle  in Fig.~\ref{fig:particle_image}(a)],
sometimes more [c.f. particle in the centre of Fig.~\ref{fig:particle_image}(a)].
We investigate the orientational dynamics of highly asymmetrical particles 
by studying aggregates of glass rods. Following \citet{Lewandowski2008}
a dilute suspension of microrods is left to evaporate in order to produce  double particles,
Fig.~\ref{fig:particle_image}(b)\obs{--(d)}.
The glass particles have an index of refraction of $1.56$ and a density of \unit[$\rho_{\rm p}$ = 2.56]{g cm$^{-3}$}.

To achieve neutral buoyancy the fluid must have the  
same density as the particles. \obs{To this end we mix}   \unit[22.2]{$\%$wt} water, \unit[4.4]{$\%$wt} glycerol, and \unit[73.4]{$\%$wt} sodium metatungstate monohydrate (Alfa Aesar GmbH). 
\obs{Density matching is achieved by titration while observing the particle under the microscope until the particle
is buoyant for several minutes. But there are additional sources of error that are difficult to control.
For example, when the fluid is  pumped through the microchannel minute changes in temperature are expected
to slightly change the fluid density. Such changes could induce small variations in the $y$-position of the particle
that we sometimes observe during the recordings.}
The suspension is highly diluted in order to avoid particle-particle interactions. The mixture has a viscosity of 
\unit[$\mu$ = 25]{mPa s} at \unit[20]{$^o$C}. We estimate the shear Reynolds number ${\rm Re}_s= \rho_{\rm f} s a^2/\mu$ as follows.  
The particle length $a$ is of the order of \unit[20]{$\upmu$m}. The shear rate $s$ is determined
by the flow speed $u_x$ which is in turn given by the flow rate $8$ $\mu$l/min and 
the cross section of the channel, \unit[2.5]{mm} $\times$ \unit[200]{$\upmu$m}.
This results in $u \approx 0.3$mm s$^{-1}$ and a shear rate of $s \approx 3 {\rm s}^{-1}$ at $60\upmu$m depth assuming
a parabolic profile in the $y$-direction, and that the channel
is much wider than deep.
The density of the fluid $\rho_{\rm f}$ equals the particle density, \unit[$\rho_{\rm p}$ = 2.56]{g cm$^{-3}$}.  
This gives ${\rm Re}_s \approx 10^{-4}$. Inertial effects are thus negligible
on the time-scale of the experiment.

The microchannel is produced using standard soft lithography. The process begins by milling a rectangular moulding form in aluminum. 
The surfaces of the mould are mechanically polished. A $10$:$1$ mixture of 
polydimethylsiloxane (PDMS) and sylgard 184 Dow Corning 
(Sigma Aldrich) is poured on the moulding form and  allowed to cure for several hours. The PDMS replica obtained after peeling off the mould is  sealed to a glass slide (thickness \unit[0.17]{mm})  by oxygen plasma bonding. This results in a rectangular channel that is  \unit[40]{mm} long, \unit[2.5]{mm} wide, and \unit[150]{$\upmu$m} deep. For some measurements, a  channel with a depth 
of \unit[200]{$\upmu$m} was used, see Table \ref{tab:1}.
The suspension is injected into the channel using thin tubing connected to a syringe pump. 
PTFE tubes  (Cole-Parmer) with outer diameters of \unit[0.76 or 1.07]{mm} are used, the former in connection with  a plastic syringe 
(\unit[1]{ml}, Terumo), the latter in connection with a glass  syringe (\unit[500]{$\upmu$l}, Hamilton).

The optical system is built around a Nikon $X$60 microscope objective (NA 1.0, WD  \unit[2]{mm}). 
The particle motion is recorded with a CCD video camera (Leica). The channel is mounted on a translation stage that 
moves the microchannel over the fixed observation microscopic objective.  
The stage is driven by a stepper motor that records the position of the stage. By moving the channel a given particle is kept within the field
of view of the objective, despite the fact that the particle is advected by the fluid through the microchannel.

A single-beam optical trap  is used to set the initial orientation and position  of a given particle. 
The optical trap makes use of the microscope objective (Fig.~\ref{fig:exp_setup}), it provides sufficient magnification to
not only visualise the particle, but also to trap it with a continuous infrared laser of wavelength \unit[1075]{nm} (\unit[10]{W}, IPG-Laser GmbH).
The most efficient way of trapping a glass rod with this setup is to direct the laser beam towards one of the two ends of the particle. 
Different orientations can be imposed  on the particle by moving the channel sufficiently quickly to cause one end 
of the particle to leave the trap, yet sufficiently slowly so that the other end of the particle is kept trapped. 

\obs{The experimental procedure described above ensures stable and reproducible experimental conditions. But a number of external factors
were difficult to control and could still influence the experimental results to a small degree: evaporation due to the presence of unsaturated air in the reservoir, the presence of air bubbles in the channel that are difficult to get rid of, and pressure drops ininside the channel  
due to the elasticity of the PDMS structure. }

We use the image-analysis algorithm employed by \citet{Einarsson2013a}. Images are acquired 
at a rate of $100$ frames per second. 
Each frame is stored as $8$-bit gray-scale bitmap with
$692 \times 520$ pixels. The pixel size is $0.21\upmu$m. 
For a given frame the image analysis proceeds in three basic steps. First static noise is reduced by
subtracting the time-averaged intensity from each frame. Then the 
boundary of the projection of the particle into the image plane is detected, and finally
an ellipse is fitted to the boundary. Details are given in Ref.~\citenum{Einarsson2013a}. 
The output defines the position and the orientation of the particle in the image.  
The centre-of-mass coordinates of the particle in the laboratory frame are determined 
using the output from the stepper motor recording the position of the stage.

\obs{The main sources of uncertainty in the image analysis are the limited resolution of the camera and diffraction. The latter
gives rise to a diffuse particle boundary. The uncertainty in the determination of the particle orientation is largest when the
short end of the rod faces the camera, i.e. for small values of $n_z$ (see Fig.~\ref{fig:exp_setup}).}

\begin{figure}[p]
\begin{overpic}[width=14cm]{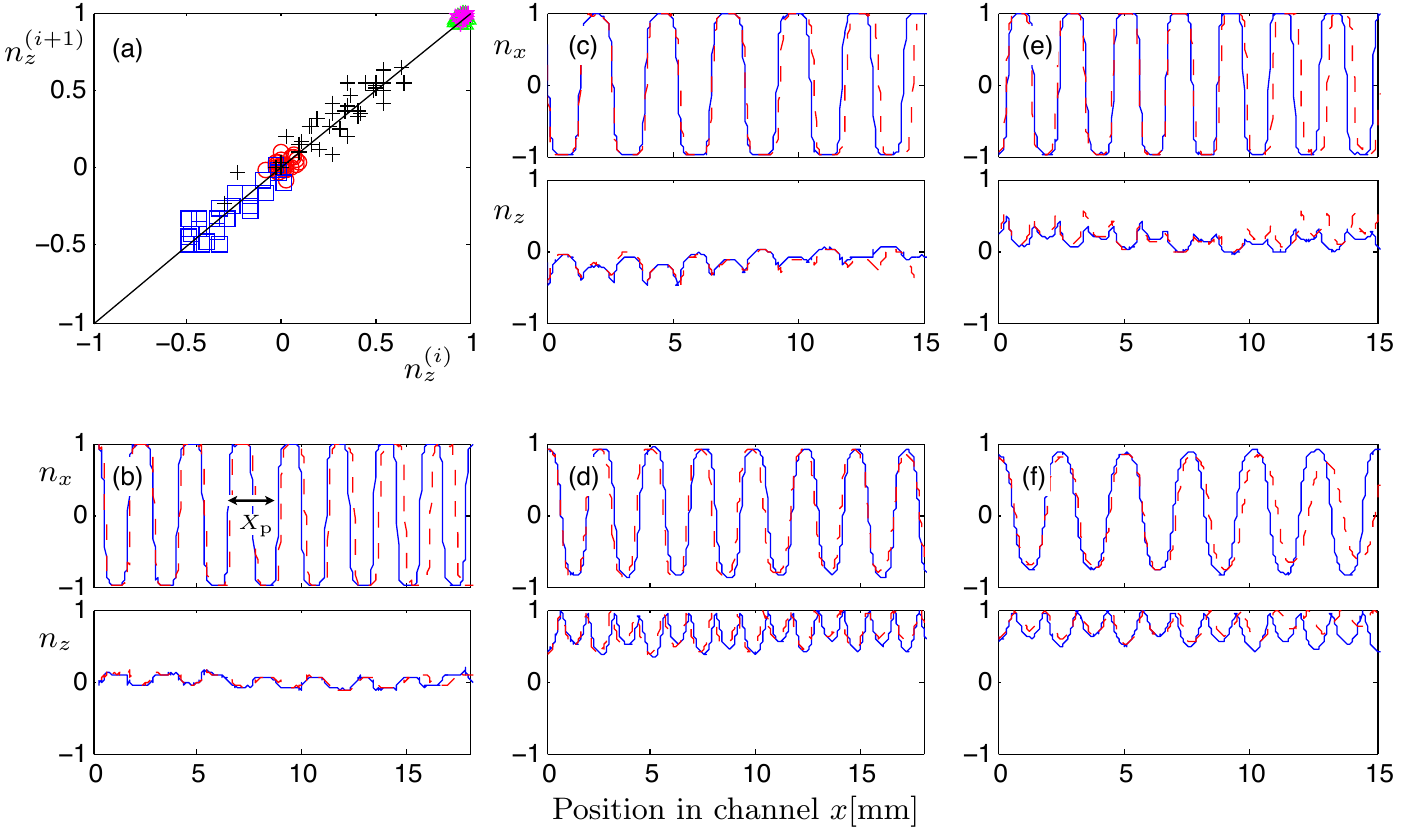}
\end{overpic}
\caption{\label{fig:particle1}
Orientational dynamics of  particle  $1$. 
(a) Dynamics of $n_z$. Here $n_z^{(i)}$ denote the values of $n_z$ at subsequent zero crossings of $n_x$, $i=1,2,3,\ldots$. The data are taken from panels (b-f). 
Red $\circ$ data from panel (b);  blue $\Box$ data from (c);  green $\triangle$ data from (d); black $ + $ data from (e); magenta $\star$ data from (f). 
Panels (b-f) show orientational dynamics as a function of c.o.m.-position $x$ in the channel, Eq. (\ref{eq:tx}). Data in different panels correspond to different initial orientations.
Solid blue and  dashed red lines represent forward and backward trajectories, respectively. The flow direction is reversed at $x=0$. 
The horizontal arrow in panel (b) indicates the period $X_{\rm p}$[mm] of the trajectory. The video-microscopy recording of the orientational dynamics shown in panel (b) can be viewed at MULTIMEDIA VIEW, that of panel (d) at 
MULTIMEDIA VIEW.}
\end{figure}

A typical experiment starts by capturing a particle with the optical trap. 
The particle is brought into the desired location in the $x$-$z$-plane by moving the channel. All particles are
started  close to one of the inlets at approximately equal $z$-distances to both side walls. We verified
that the $z$-position remains centred, with an 
error typically at most one particle length.
This implies that the shear in the $z$-direction remains very small (the channel is much wider than deep).
The $y$-coordinate thus corresponds to the shear direction, and the $z$-coordinate is the vorticity direction.
\obs{The particle is brought to the desired initial orientation 
as described above and the centre-of-mass of the particle is placed at a depth of $60$ mm. The precision in determining the initial depth 
is determine by the depth-of-field of the microscope. It is estimated to be of the order of one particle length.}
\obs{Then the particle is} released
to follow the flow in the microchannel. 
We then invert the pressure gradient so that the particle is advected back in the opposite $x$-direction. 
For each orientational trajectory we record both forward and backward dynamics.
Since Stokes' equation is invariant under simultaneous pressure inversion and time reversal, the backward dynamics must exactly
retrace the forward dynamics unless irreversible effects such as inertia or thermal noise affect the dynamics.
Examples for the resulting video-microscopy recordings of the orientational dynamics in the $x$-$z$-plane can be viewed 
via the MULTIMEDIA VIEW links in Figs.~\ref{fig:particle1} and \ref{fig:particle3}.

\begin{figure}[p]
\begin{overpic}[width=14cm]{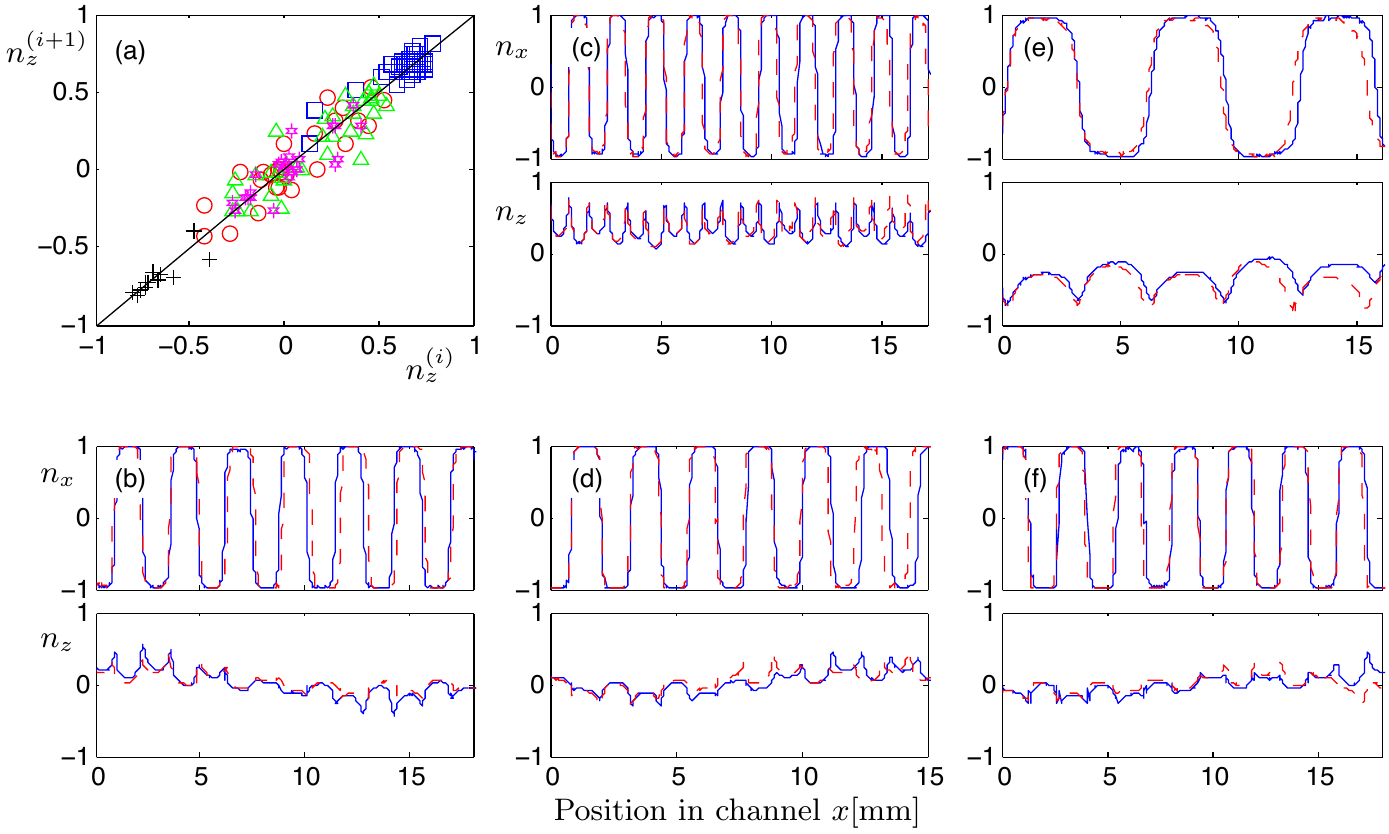}
\end{overpic}
\caption{\label{fig:particle2} Orientational dynamics of  particle  $2$. See caption of Fig.~\ref{fig:particle1} for details. }
\end{figure}

For a given particle this procedure is repeated many times to obtain orientational trajectories with different initial orientations.
We record the length of the projection of the particle into the $x$-$z$-plane as a function of time. We estimate the particle length 
using the procedure described in 
Ref.~\citenum{Einarsson2013a}. 
Once the particle length is known we can extract the components
of the unit vector $\ve n$ determining the orientation of the particle, as a function of time. 

We plot the orientation not as a function of time but as a function of distance that the centre-of-mass of the particle
has traveled through the channel, advected by the flow:
\begin{equation}
\label{eq:tx}
x(t) = \int_0^t\!\!{\rm d}t' u_x(t')\,.
\end{equation}
Here $u_x(t)$ is the instantaneous flow velocity at time $t$. This transformation simplifies the analysis because it accounts for
the fact that the shear-rate is time-dependent: the flow velocity changes when the pressure is reversed, and
in order to avoid inertial effects these reversals must be performed slowly. The invariance of Stokes equation under
time and pressure reversal implies $x(t) = -x(-t)$. We overlay forward and backwards dynamics by plotting the orientation as a function of centre-of-mass position.

\section{Experimental results}
\label{sec:results}
\begin{figure}[t]
\begin{overpic}[width=14cm]{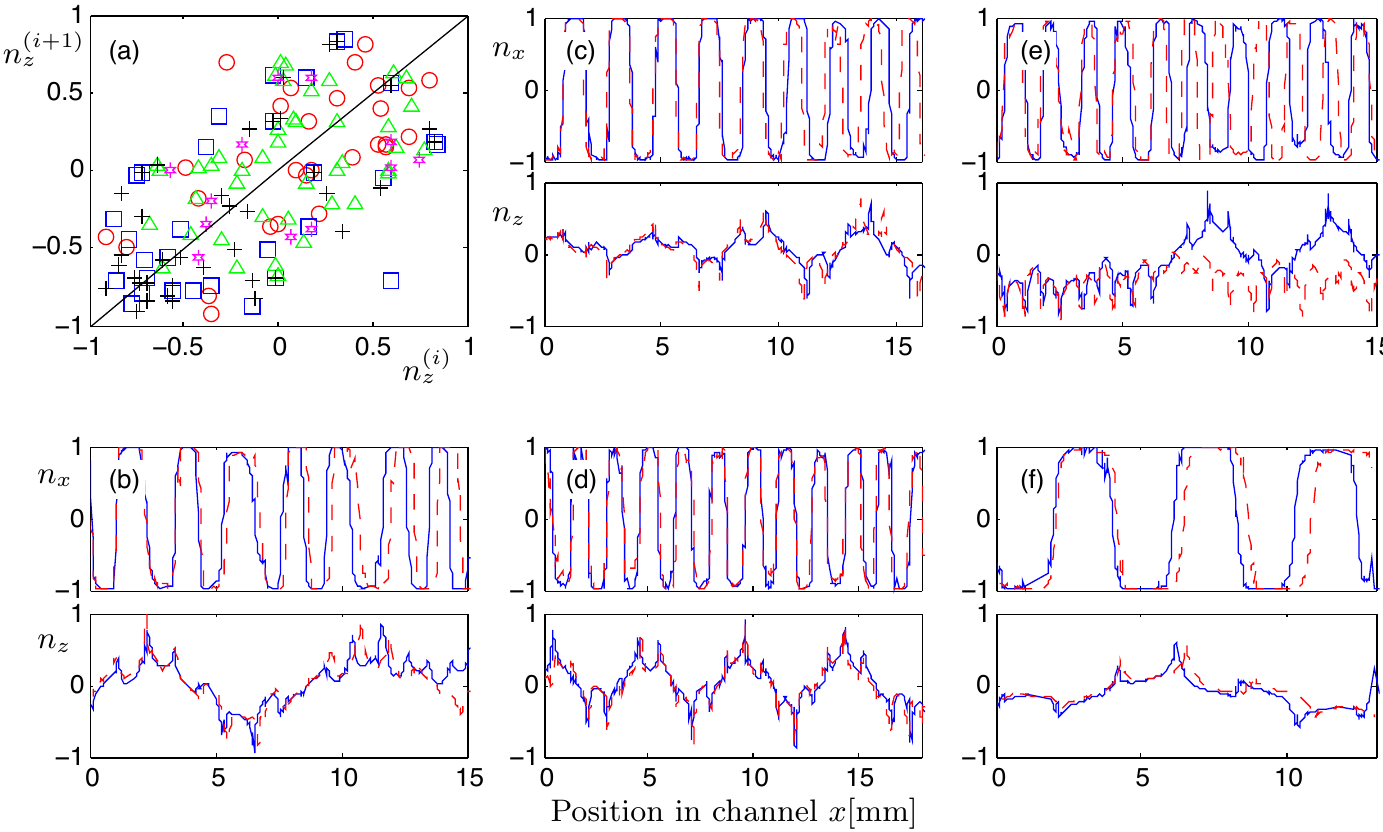}
\end{overpic}
\caption{\label{fig:particle3} Orientational dynamics of  particle  $3$. See caption of Fig.~\ref{fig:particle1} for details. The video-microscopy recording of the orientational dynamics shown in panel (d) can be viewed via MULTIMEDIA VIEW}
\end{figure}

Figures~\ref{fig:particle1}, \ref{fig:particle2}, and \ref{fig:particle3} show orientational dynamics of three different particles.
For each particle five different orientational trajectories are shown, corresponding to different initial orientations [panels (b) to (f)].
In all three Figures we show trajectories of the $x$- and $z$-components of the unit vector $\ve n$ that points along the major axis of the particle. Here
$n_x$ is the component  in the flow direction, and $n_z$ is the component in the vorticity direction. The third component of $\ve n$ is determined by normalisation, 
$|\ve n|=1$.

In panel (a) of each Figure we summarise the orientational dynamics by 
recording the values of $n_z$ whenever $n_x=0$. We denote the resulting sequence of consecutive $n_z$-values by $n_z^{(i)}$, $i=1,2,3,\ldots$. For an axisymmetric particle Jeffery's equation predicts that $n_z^{(i+1)}=n_z^{(i)}$, shown as the solid line along the diagonal in panel (a).

As mentioned above, the particle is first advected along a stream line
of the pressure-driven flow in the channel. We then 
invert the pressure gradient so that the centre-of-mass of
the  particle is advected back to where it came from.
For each orientational trajectory we show the \lq forward dynamics\rq{} (blue solid line), going from right-to-left in the Figure. 
After the {reversal} follows the \lq backward dynamics\rq{} (red dashed line). Since Stokes' equation is invariant under simultaneous pressure inversion and time reversal, 
the backward orientational dynamics must  exactly retrace the forward dynamics unless irreversible effects due to  inertia or thermal noise affect the dynamics
noticeably on the time-scale of the experiment.

Consider first the trajectories shown in Fig.~\ref{fig:particle1}, corresponding to particle $1$. 
Panels (b) to (f) show orientational trajectories of $n_x$ and $n_z$ for different initial orientations. In all cases the backward dynamics retraces the forward dynamics very well.  
This shows that neither inertial forces nor rotational diffusion affect the orientational dynamics. 
We attribute the small dephasing visible in each panel
to a small density mismatch \obs{(discussed in Section~\ref{sec:methods})}, causing the particle to sink (or float), changing the shear rate it experiences.
Apart from this slight dephasing all orientational trajectories are fairly 
periodic. For a given $n_x$-trajectory the relative variation of the centre-of-mass distance between two consecutive
$n_x\!=\!0$-events (\lq half-period\rq{}) is of the order  of $10$\%.
Between different panels we observe variations in the period $X_{\rm p}$, between \unit[2.1]{mm} and \unit[2.6]{mm}, \obs{caused by the uncertainty in the $y$-position mentioned above.}  Panel (a) indicates that $n_z^{(i+1)}$ is approximately equal
to $n_z^{(i)}$.

Fig.~\ref{fig:particle2} shows the orientational dynamics of particle $2$
for different initial orientations [panels (b) to (f)]. In all cases the backward dynamics retraces the forward dynamics well, at least for a few millimetres. As for particle $1$ the trajectories show a slight dephasing within each panel.
Different panels show quite different periods $X_{\rm p}$, ranging between $1.7$mm and $5.9$mm.  In panels (d) and (f) we see that the amplitude of $n_z$  changes aperiodically. This is also apparent from panel (a) exhibiting a somewhat wider scatter around the diagonal than panel (a) in Fig.~\ref{fig:particle1}.

Fig.~\ref{fig:particle3} shows the orientational dynamics of particle $3$, an asymmetric double particle. For all initial orientations the backward dynamics retraces the forward dynamics fairly well, with the exception of the trajectories shown in panel (e). As for particles $1$ and $2$ we observe a slight dephasing within each panel. The dynamics of $n_x$ is quite periodic, but again the periods $X_{\rm p}$ vary from panel to panel, ranging  from \unit[1.4]{mm}
 to \unit[4]{mm}. In all cases $n_z$ shows distinct aperiodicity. Panel (a) exhibits a larger scatter around the diagonal than panel (a) in Fig.~\ref{fig:particle2}.

\begin{table}[t]
 \begin{tabular}{llllllll}
  \hline\hline
   Particle&Particle\hspace*{3mm}       & Particle &  Channel\hspace*{3mm}             &Flow rate      & Figure \\
   number&type$^a$& length($\upmu$m)$^b$\hspace*{3mm}   &  depth ($\upmu$m)\hspace*{3mm}    &($\upmu$l/min)\hspace*{3mm}    &     & \\\hline
   Particle $1$ &single & 20.5 & 200 & 8   &  Fig.~\ref{fig:particle1}\\
   Particle $2$ &single & 24.3 & 150 & 5   &  Fig.~\ref{fig:particle2}\\
   Particle $3$ &double$^c$ & \obs{28.7} & 200 & 8 &  Fig.~\ref{fig:particle3}\\
   Particle $4$ &single & 20.1 & 150 & 7.5 & Fig.~S1$^d$ \\
   Particle $5$ &single & 29.5 & 200 & 8    & Fig.~S2\\
   Particle $6$ &single & 17   &150 & 5      & Fig.~S3 \\
   Particle $7$ &single & 25   & 150 & 5     & Fig.~S4 \\
   Particle $8$ &single & 26.5 & 150 & 5    & Fig.~S5 \\
   Particle $9$ &single & 22.5 & 150 & 7.5 & Fig.~S6 \\
   Particle $10$&single & 25.5 & 150 & 5   & Fig.~S7 \\
   \obs{Particle $11$}&double$^e$ & 26 & 200 & 8   & Fig.~S8 \\
   \obs{Particle $12$}&double$^f$ & 17 & 200 & 8   & Fig.~S9 \\

\hline\hline
  \end{tabular}
  \caption{Description of particles and other experimental parameters. \mbox{}$^a$Single or double particle. \mbox{}$^b$Particle length as extracted from image-analysis algorithm, see Section \ref{sec:methods}. We estimate
the error to be of the order of $1\upmu$m.  \obs{\mbox{}$^c$This particle is shown in Fig.~\ref{fig:particle_image}(b). The lengths of the two glass rods
are 26.4 and 17.3$\upmu$m.}
  \mbox{}$^d$These Figures are in the Supplementary Online Material \cite{supp}.
\obs{\mbox{}$^e$This particle is shown in  Fig.~\ref{fig:particle_image}(c). The lengths of the two glass rods are 26 and 16.5$\upmu$m. 
\mbox{}$^f$This particle is shown in  Fig.~\ref{fig:particle_image}(d). 
The lengths of the two glass rods are both 16$\upmu$m. }
}
 \label{tab:1}
\end{table}
The orientational dynamics of particles $1$, $2$, and $3$ was obtained under slightly different experimental conditions. These are summarised in Table~\ref{tab:1},
as well as the particle properties. Table~\ref{tab:1} also gives information about other particles for which we obtained precise orientational dynamics. The corresponding Figures are found in the Supplementary Online Material \cite{supp}.

Video-microscopy recordings corresponding to the data shown in 
Figs.~\ref{fig:particle1}(b), (d), and \ref{fig:particle3}(d) 
can be viewed via the MULTIMEDIA VIEW links in the corresponding figure
captions.

\section{Discussion}
\label{sec:discussion}
The results summarised in Figs.~\ref{fig:particle1} to \ref{fig:particle3} show the orientational dynamics of single and double glass rods.
In general the particles  are not perfectly axisymmetric (as seen in Fig.~\ref{fig:particle_image}), and therefore do not follow perfect Jeffery orbits. In this section we relate our experimental results to the theoretical predictions valid for ellipsoidal particles\cite{Jeffery1922,Hinch1979}. The particles in our experiment  do not satisfy the mirror symmetries assumed in this theory, but it is plausible that the effects of breaking axisymmetry  predicted by this theory apply at least qualitatively to our glass rods.

The equation of motion for the orientational dynamics of an ellipsoidal particle\cite{Jeffery1922,Hinch1979} can be cast in the form\cite{Einarsson2013b}
\begin{subequations}
\label{eq:jeffery}
\begin{align}
\dot{\ve n}&={\ma O}\ve n +\Lambda\, \big({\ma S}\ve n - (\ve n\cdot{\ma S}\ve n)\ve n\big)
+ \frac{K(1-\Lambda^2)}{K\Lambda-1}(\ve n\cdot{\ma S}\ve p)\ve p\,,\\
\dot{\ve p}&={\ma O}\ve p +K \,\big({\ma S}\ve p  - (\ve p\cdot{\ma S}\ve p)\ve p\big) + \frac{\Lambda(1-K^2)}{K\Lambda-1}(\ve n\cdot{\ma S}\ve p)\ve n\,.
\end{align}
\end{subequations}
Following the convention outlined 
in Section \ref{sec:methods}, $\ve n$ is a unit vector that points along the major axis of the ellipsoidal particle. The  unit vector  $\ve p$ 
is orthogonal to $\ve n$, directed along the particle axis corresponding
to the length $b$ used in the definition of the aspect ratio $\kappa$
(defined in the Introduction).
The geometry of the ellipsoid is characterised by the two
shape parameters $\Lambda$ and $K$ that are defined in the Introduction, Section \ref{sec:intro}.
The matrices $\ma S$ and $\ma O$ are the symmetric and anti-symmetric parts of $\ma A$, the matrix  of fluid-velocity gradients. In our case this matrix takes the form
\begin{equation}
\ma A=
  \left[ {\begin{array}{cccc}
   0 & s & 0 \\
   0 & 0 & 0 \\
   0 & 0 & 0 \\
  \end{array} } \right]\,,
\end{equation} 
where $s$ is the shear strength (Fig.~\ref{fig:exp_setup}).

Eqs. (\ref{eq:jeffery}) are symmetric under the simultaneous exchange of $\ve n$ and $\ve p$ as well as $\Lambda$ and $K$: describing the motion of the same particle using
a different coordinate system within the particle must result in the 
same dynamics.
Note also that the non-linear coupling between $\ve n$ and $\ve p$  involves the strain $\ma S$ only. This coupling maintains 
orthogonality of the two vectors $\ve n$ and $\ve p$.
 The anti-symmetric part $\ma O$
just causes a solid-body rotation. For axisymmetric particles, $K=0$, 
so that the tumbling of $\ve n$ becomes independent of
the dynamics of $\ve p$ (but not vice versa). The resulting equation for $\ve n$ has infinitely many degenerate periodic solutions, the Jeffery orbits\cite{Jeffery1922}. The dynamics of $\ve p$ describes how the particle spins around its symmetry axis.

When $K\neq 0$ (and $\Lambda \neq 0$), no general closed-form solutions 
of Eqs.~(\ref{eq:jeffery}) are known. 
It is convenient to represent the numerical solutions of Eq.~(\ref{eq:jeffery}) in terms of a Poincar\'{e} surface-of-section\cite{Strogatz}, recording the locations at which the dynamics intersects a surface in the phase space of Eq.~(\ref{eq:jeffery}). This section is constructed as follows.
\citet{Hinch1979} have shown
that the vector $\ve n$ rotates around the vorticity at a positive angular velocity so that one can reduce the dimensionality of the problem by parametrising
the orientational dynamics in terms of the corresponding angle.
A suitable condition\cite{Hinch1979,Yarin1997}  defining the surface-of-section  is that $\ve n$ is perpendicular to the flow direction, $n_x=0$.
Following Yarin {\em et al.}\cite{Yarin1997} we take the coordinates in the surface-of-section to be $\psi$
and $n_z$, where
$\psi$ is the Euler angle parametrising the spin of the particle around the axis $\ve n$, and $n_z$ is the $z$-component of $\ve n$,  
the cosine of the angle $\theta$ between $\ve n$ and vorticity (Fig.~\ref{fig:exp_setup}). 
When $n_x=0$  we record 
the coordinates $(\psi, n_z)$. 
To define
the Euler angles ($\theta,\phi,\psi$) we use the convention of Goldstein\cite{Goldstein} and express $\ve n$ and $\ve p$ as
\begin{equation}
\ve n = 
\left [\begin{array}{l}\phantom{-}\sin\theta\sin\phi \\
                       -\sin\theta\cos\phi \\
                       \phantom{-}\cos\theta\end{array}\right]\,,\quad\mbox{and}\quad 
\ve p = \left [\begin{array}{l}-\sin\psi\cos\phi-\cos\theta\sin\phi\cos\psi \\
                       -\sin\psi\sin\phi + \cos\theta\cos\phi\cos\psi \\
                       \phantom{-}\cos\psi\sin\theta\end{array}\right]\,.
\end{equation}
So $\theta$
is the polar angle depicted in Fig.~\ref{fig:exp_setup}. The angle $\phi$ is referred to as the \lq precession angle\rq{}. This angle measures the direction of the projection of  $\ve n$ into  the flow-shear plane.
Eqs.~(\ref{eq:jeffery}) correspond to Eqs.~(2.1) and (2.2) in Ref.~\citenum{Yarin1997}, setting $s=-1$ and defining the aspect ratios $\lambda$ and $\kappa$
in terms
of the particle axes\cite{Yarin1997} as follows: $a_x=\lambda a_z$ and $a_y = \kappa a_z$.

In the experimental time series shown in Figs.~\ref{fig:particle1} to \ref{fig:particle3},  instances where $n_x\! =\! 0$ correspond  to 
peaks in the oscillation in $n_z$. The $n_z$-coordinate in the surface-of-section is therefore easily read off from the experimentally observed time series. The angle $\psi$, by contrast, cannot be measured in our experiment because we cannot track how the particles spin around $\ve n$.

\begin{figure}[p]
\begin{overpic}[width=14cm]{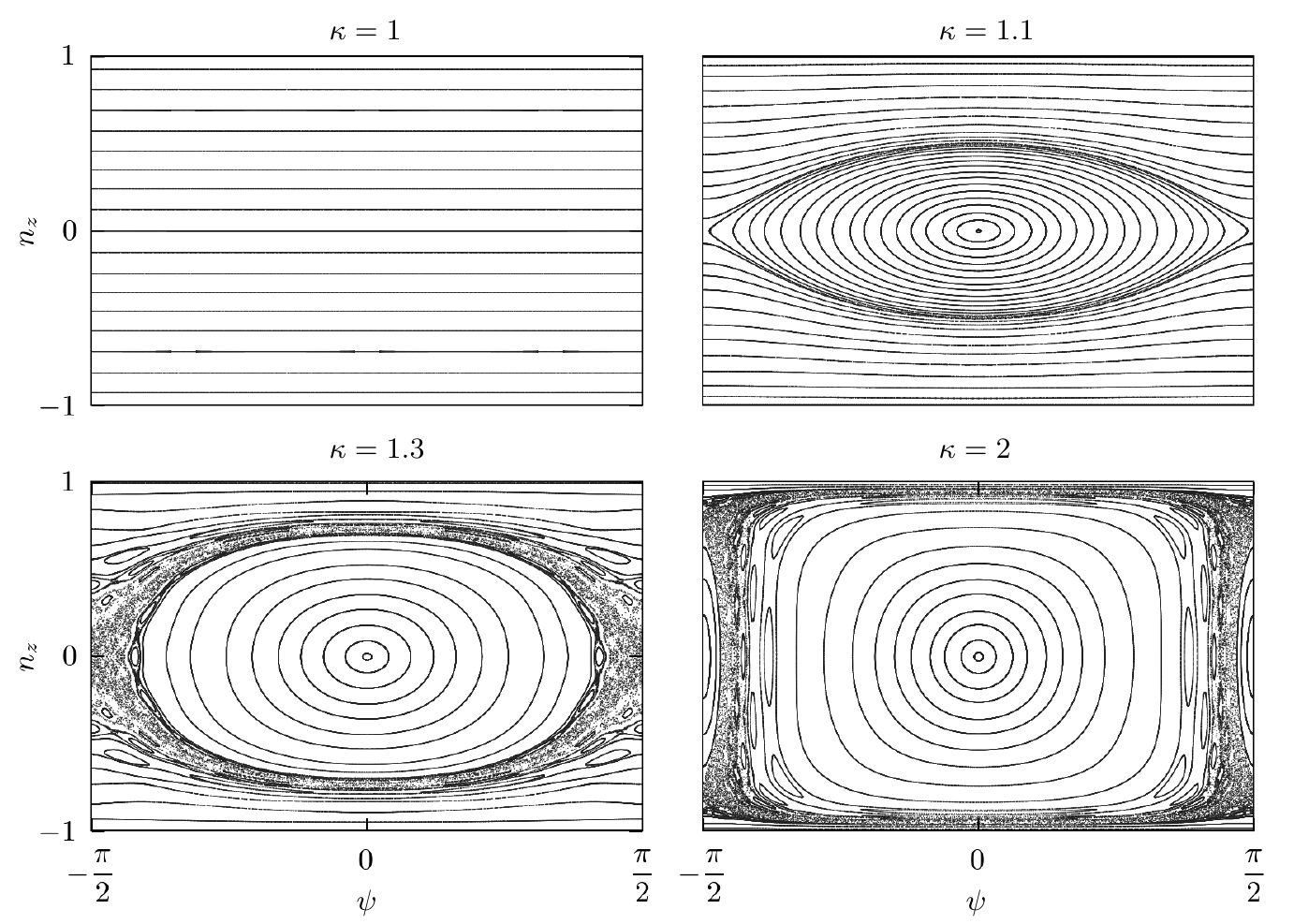}
\end{overpic}
\caption{\label{fig:poincares} Poincar\'e{} surfaces-of-section for $\Lambda=12/13$ and different values of $\kappa$  
[$K=(\kappa^2-1)/(\kappa^2+1)$]. The angle $\psi$ is defined up to $(\psi+\pi/2)\mbox{mod}\pi$.}
\end{figure}

Four different surfaces-of-section are shown in Fig.~\ref{fig:poincares}, obtained by numerical integration of Eqs.~(\ref{eq:jeffery}) for a large number of different initial orientations,
and plotting the sequence $[\psi^{(i)},n_z^{(i)}]$ of $(\psi,n_z)$ evaluated at consecutive zero crossings of $n_x$, labeled
by $i = 1,2,3,\ldots$. Similar sections are found in Ref.~\citenum{Yarin1997}. The 
map that gives $[\psi^{(i+1)},n_z^{(i+1)}]$ in terms of $[\psi^{(i)},n_z^{(i)}]$ is called the Poincar\'e{} map.

Fig.~\ref{fig:poincares}(a) depicts the orientational dynamics of an axisymmetric particle, $K=0$. The coordinate $n_z$ is a conserved quantity on the surface-of-section, Jeffery orbits appear as horizontal lines in Fig.~\ref{fig:poincares}(a), one-parameter families parametrised by $\psi$. In the literature Jeffery orbits are commonly identified by their orbit constant $C$. It is given by the value of $\tan\theta$ at $n_x=0$ (see for example Eq.~(3) in Ref.~\citenum{Hinch1979}). In this paper we characterise Jeffery orbits by $n_z=\cos\theta=1/(1+C^2)$ on the surface-of-section ($n_x=0$). Fig.~\ref{fig:poincares}(a) illustrates that the orientational dynamics depends on the initial orientation, determining the value of $n_z$.
We remark that the periods of $\ve n$ and $\psi$ are not in general commensurate for $K=0$. But  the tumbling of $\ve n$ is independent of that of the angle $\psi$ and thus periodic for axisymmetric particles.

Fig.~ \ref{fig:poincares}(b) shows the orientational dynamics of a weakly asymmetric ellipsoidal 
particle ($K \approx 0.095$ and $\Lambda=12/13$). We see that Jeffery orbits
with $n_z \approx \obs{\pm} 1$ remain almost unchanged. 
But there are substantial changes for smaller values of
$|n_z|$, compared with the surface-of-section for $K=0$. 
We see that $n_z$ ceases to be a constant of motion,
doubly-periodic orientational dynamics results. The most substantial
changes occur near $n_z=0$. The $n_z\!=\!0\,$-Jeffery orbit 
is replaced by two fixed points at $(0,0)$ and $(\pm\pi/2,0)$ on the surface-of-section.

This fact and the surface-of-section patterns in the vicinity of these points follow from the time-reversal symmetry
of Eqs.~(\ref{eq:jeffery}). The general principle is explained in Section 6.6 of Ref.~\citenum{Strogatz}. See also Ref. \citenum{Politi1986}.
The invariance of Stokes equation referred to in Section \ref{sec:methods} implies that Eqs. (\ref{eq:jeffery}) are invariant
under
\begin{equation}
\label{eq:symmetry}
t \to -t\,,\quad n_x \to -n_x\,,\quad\mbox{and}\quad p_x\to -p_x\,.
\end{equation}
The fixed point $(0,0)$ is mapped onto itself under this transformation. It follows that
the dynamics in its immediate neighbourhood can neither be expanding nor contracting. 
In other words the determinant describing the linearised motion in the vicinity of this fixed point,
\begin{equation}
\det \left [ 
\begin{array}{ll}
\frac{\partial \psi^{(i+1)}}{\partial \psi^{(i)}} & \frac{\partial \psi^{(i+1)}}{\partial n_z^{(i)} }\\
 \frac{\partial n_z^{(i+1)}}{\partial \psi^{(i)}}& \frac{\partial n_z^{(i+1)}}{\partial n_z^{(i)}}\end{array}\right ] \,,
\end{equation}
must be of unit modulus despite the fact that the dynamics (\ref{eq:jeffery}) is dissipative.
We find numerically that the eigenvalues are $\exp(\pm {\rm i}\sigma)$. The point $(0,0)$ is thus an elliptic fixed point, surrounded
by a one-parameter family of closed orbits that appear as concentric closed curves, much like so-called \lq tori\rq{} in so-called \lq Hamiltonian\rq{} systems with area-preserving phase-space dynamics\cite{Lichtenberg}.
For near-axisymmetric particles \citet{Hinch1979} analysed the corresponding orbits by multiple-scales analysis.
As these orbits rotate around the elliptic point, the value of
$n_z$ changes sign. This doubly-periodic motion may be quasi-periodic or  periodic, depending on whether the two frequencies are incommensurate
or not (corresponding to irrational or rational winding numbers, respectively).
\begin{figure}[p]
\begin{overpic}[width=9cm]{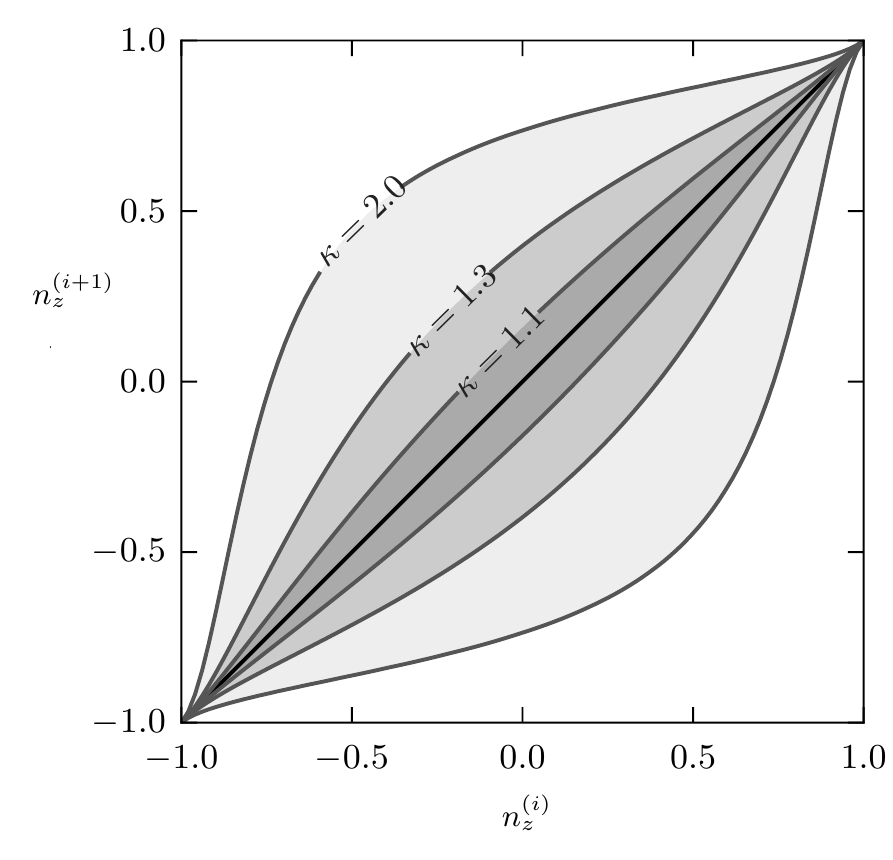}
\end{overpic}
\caption{\label{fig:max_nz} Range of Poincar\'e{} map for 
$\Lambda=12/13$ and different values of $\kappa$ corresponding
to the values used in Fig.~\ref{fig:poincares}, $K=(\kappa^2-1)/(\kappa^2+1)$.
The data shown was obtained by 
\obs{by iterating the Poincar\'e{} map once for
randomly chosen initial conditions. The initial conditions were determined by sampling $\psi$ uniformly over the surface-of-section for given values of $n_z$.}
}
\end{figure}

The point ($\pm \pi/2,0$) is a hyperbolic fixed point.
\obs{We find numerically that the eigenvalues are  real 
and opposite}, $\gamma$ and $\gamma^{-1}$.
The fact that the $n_z\!=\!0\,$-orbit is destroyed upon infinitesimal perturbation and replaced by a discrete set
of fixed points of alternating stability is typical for Hamiltonian systems\cite{Lichtenberg}. The mechanism in our dissipative system is analogous,
a consequence of the symmetry (\ref{eq:symmetry}).

For larger asymmetries chaotic orientational dynamics occurs, seen as a region with a stochastic scatter of points
in Fig.~\ref{fig:poincares}(c). Almost all Jeffery orbits are modified, only those with $n_z$ close to $\pm 1$ remain.
Whether   the orientational dynamics is periodic, quasi-periodic, or chaotic  depends upon the initial condition on the surface-of-section.

Fig.~\ref{fig:poincares}(d) shows the orientational dynamics for ellipsoidal particles with $K = 3/5$ and $\Lambda=12/13$.
This value of $K$ is similar the corresponding value for the double particle $3$ (Figs.~\ref{fig:particle_image}(b) and \ref{fig:particle3}),
but this particle has $\Lambda \approx 40/41$. We do not seek quantitative correspondence between these parameters because the particles used in our experiment are not ellipsoidal. This prevents us from drawing quantitative conclusions, but still allows for a qualitative comparison between theory and experiment.
The orientational dynamics displayed in Fig.~\ref{fig:poincares}(d) is either on tori or chaotic. 
Motion on tori  can occur with large amplitudes, so that $n_z$ changes from $n_z \approx -1$ to $n_z \approx 1$. 

How are these observations reflected in the experimental time series shown in Figs.~\ref{fig:particle1} to \ref{fig:particle3}?
Since we cannot measure the angle $\psi$ in our experiments, we concentrate on the dynamics of $n_z$.
Fig.~\ref{fig:max_nz} shows the range of changes of $n_z$ in one iteration of the Poincar\'e{} map  for particles with different degrees of asymmetry, determined by numerically recording the changes $n_z^{(i+1)}-n_z^{(i)}$ along orientational trajectories. The diagonal corresponds to symmetric particles where $n_z^{(i+1)} = n_z^{(i)}$. The larger the asymmetry, the larger is the range of   $n_z^{(i+1)}-n_z^{(i)}$ that may occur, reflecting the properties of the surfaces-of-section shown in Fig.~\ref{fig:poincares}.

Fig.~\ref{fig:max_nz}  can be directly compared with Figs.~\ref{fig:particle1}(a) to \ref{fig:particle3}(a). Our experimental results show that the Poincar\'e{} map may scatter significantly around the diagonal. The above discussion explains that this is a consequence of doubly-periodic and possibly 
chaotic orientational dynamics of asymmetric particles. The range of scatter differs between different particles, a consequence of different degrees of asymmetry. We see that the scatter is largest for the double particle $3$, with $K \approx 3/5$
[Fig.~\ref{fig:particle3}(a)].
Particle $1$, by contrast, shows only negligible scatter around the diagonal [Fig.~\ref{fig:particle1}(a)]. We infer that this particle is highly symmetric, 
$K$ is very small.
The data shown in Fig.~\ref{fig:particle1}  are consistent with the conclusion that particle $1$ follows Jeffery orbits, $n_z$ remains approximately constant
in the surface-of-section. But we remark that the shape of the $n_z\approx 0$-orbit does not look like a Jeffery orbit for an axisymmetric particle. 
We cannot exclude that this is due
to possible non-ellipsoidal deviations from axisymmetry. The surface-of-section dynamics is most sensitive to such shape perturbations
near $n_z=0$. 
A more likely explanation is that the shape of the trajectory is a consequence of  systematic (and reproducible) tracking
errors due to diffraction and finite pixel size. \obs{These errors are largest for small values of $n_z$, as discussed
in Section~\ref{sec:methods}.}

Particle $2$ is also a single glass rod, but it shows a somewhat larger range of scatter around the diagonal [Fig. \ref{fig:particle2}(a)].
We attribute this to a more substantial breaking of axisymmetry at the tips of the particle [as seen for instance in Fig.~\ref{fig:particle_image}(a)]. Particle $2$ shows fairly periodic motion for $n_z \approx 1$ and distinctly doubly-periodic orientational dynamics for small values of $n_z$. This confirms that the initial orientation determines whether the particle tumbles 
periodically or aperiodically.  Note also  that $n_z$ changes sign along the trajectories that remain near $n_z=0$ [Fig. \ref{fig:particle2}(b), (d), and (f)].
 All of these  observations are qualitatively consistent with the surfaces-of-section shown in 
Fig. \ref{fig:poincares}(b) and \ref{fig:poincares}(c).

For particle $3$ the value of $n_z$ changes sign for all orientational trajectories shown.
This is consistent with the fact that the Fig. \ref{fig:poincares}(d)  shows
predominantly this type of motion.  Fig.~\ref{fig:particle3}(e) is consistent with chaotic orientational motion in the stochastic layer around the elliptic island. The surfaces-of-section in Fig.~\ref{fig:poincares} show that there are two types of doubly-periodic motion: either $n_z$ has the same sign, or its sign changes periodically.  The trajectory in Fig.~\ref{fig:particle3}(e) exhibits both behaviours, indicating chaotic dynamics.
Now consider the trajectory shown in Fig.~\ref{fig:particle3}(c). 
It is not periodic or doubly-periodic,  but the reversal works well 
(at least initially). But we cannot conclude that the dynamics is consistent with surfaces-of-section discussed above. The $n_z$-values 
at $n_x=0$ in Fig.~\ref{fig:particle3}(c) change from approximately $0.25$ to $0.75$ in modulus. 
Explaining this behaviour in terms of chaotic dynamics on the surfaces-of-section
requires a large stochastic region, larger than the one shown in 
Fig. \ref{fig:poincares}(d). 
This might mean that the trajectory shown in panel (d) is a doubly-periodic piece of a chaotic trajectory that may show different
behaviours at larger times. But  to determine whether these behaviours can be explained by chaotic dynamics would require to derive and numerically integrate the 
orientational equations of motion for the precise shape of the particle, and to experimentally record the angle $\psi$. Since the current experimental
setup does not allow to reliably extract how the angle $\psi$ changes, 
we plan to perform experiments with small triangular platelets 
that will allow to record the angle $\psi$.

The winding number of the trajectory shown in Fig.~\ref{fig:particle3}(d) is roughly $7$ corresponding to seven $n_x\!=\!0\,$-crossings 
while the trajectory winds approximately once round the elliptic fixed point in the centre of the surface-of-section. The winding number
of the trajectory in Fig.~\ref{fig:poincares}(e) is roughly $4$ corresponding to four $n_x=0$-crossings while the surface winds once around the elliptic fixed point. 
These observations are in qualitative agreement with the fact that the winding numbers of  tori winding around the elliptic fixed points increase as
the distance from that point increases (corresponding to larger maximal values of $n_z$).

Additional results for seven more particles are shown in the Supplementary Online Material \cite{supp}, Figs.~S1 to S7. In general the results shown in these Figures
support the observations and qualitative conclusions summarised above. But the trajectory shown in Fig.~S7 is difficult to reconcile with the surfaces-of-section
shown in Fig.~\ref{fig:poincares}. Fig.~S7(c) shows an orbit near $n_z=0$ where $n_z$ appears not to change sign.
But panels (d) and (f) in Fig.~S7 show distinct sign changes, not consistent with the surfaces-of-section shown in Fig.~\ref{fig:poincares}. 

It was pointed out in Section \ref{sec:results} that the periods $X_{\rm p}$ observed in the orientational trajectories can differ substantially between different trajectories
of the same particle. Consider for instance particle $1$ (Fig.~\ref{fig:particle1}). 
For an axisymmetric ellipsoidal particle ($K=0$) the Jeffery period (time units) is given by\cite{Jeffery1922}
 \begin{equation}
 \label{eq:Tp}
 T_{\rm p} = \pi \frac{\lambda^2+1}{s\lambda}\,.
 \end{equation}
For $K=0$ this is twice the return time to the surface-of-section. For  $K\neq 0$ the return
 time depends upon the starting position on the surface-of-section, but for small values of $K$ the deviations
 from (\ref{eq:Tp}) are small, of the order of $K$. For the nearly axisymmetric particle $1$ we can use Eq. (\ref{eq:Tp}) to estimate the period $X_{\rm p}$
in Fig.~\ref{fig:particle1}. Using the parameters from Table \ref{tab:1} and
assuming that the particle was located at a depth of \unit[60]{$\upmu$m},
Eq. (\ref{eq:Tp}) gives $X_{\rm p} \approx$~\unit[2.3]{mm}. 
This is consistent with the range of periods observed in Fig.~\ref{fig:particle1}(b) to (f),
namely \unit[2.1]{mm} to \unit[2.6]{mm}. 
We infer that the precision in determining the depth at which particle $1$ moves through the channel is of the order
of one particle length. The variations in periods observed in Figs.~\ref{fig:particle2} and \ref{fig:particle3} indicate that the actual depths vary more
between different panels than in Fig. \ref{fig:particle1}.

\section{Conclusions}
\label{sec:conclusions}
Theory and numerical simulations predict that  the orientational dynamics of small neutrally buoyant particles in  a shear flow is very
sensitive to breaking of axisymmetry\cite{Gierszewski1978,Hinch1979,Yarin1997}. 
Axisymmetric particles tumble periodically on Jeffery orbits, but when the symmetry is broken, the Jeffery orbits are modified. Depending on the initial orientation, periodic, doubly-periodic, or chaotic tumbling may result. 

\obs{In order to experimentally verify these predictions it is necessary to use particles of well-defined shape, to ensure that inertial effects and rotational diffusion are negligible, and to compare different orientational trajectories (corresponding
to different initial orientations) of the same particle.}

In this paper we describe measurements of the orientational motion of small glass rods suspended in a micro-channel shear flow.
\obs{The measurements are precise enough to allow for a definite comparison with theory\cite{Hinch1979,Yarin1997}.  
First, the glass rods have highly symmetric circular cross sections. Slight imperfections at the ends break the axisymmetry
weakly for some particles [Fig.~\ref{fig:particle_image}(a)]. We also use strongly triaxial particles designed for our purpose 
by joining two glass rods.  Second, by reverting the pressure-driven flow we observe that the orientation retraces its trajectory for many periods. This means that neither inertial effects nor rotational diffusion matter on this time scale.  Third by means of an optical trap we change
the orientation of the particle, making it possible to measure different orientational trajectories for the same particle.}

Our results \obs{confirm the theoretical predictions\cite{Hinch1979,Yarin1997}. We observe periodic and doubly-periodic tumbling, 
and how the nature of the orientational dynamics depends upon the initial orientation, and on the degree to which axisymmetry is broken. 
Our measurements are consistent with the features of the surfaces-of-section shown in Figs.~\ref{fig:poincares} and \ref{fig:max_nz}.}

In the future we plan to experimentally map out Poincar\'{e} surfaces-of-section (Fig.~\ref{fig:poincares}). With the present particles this is not possible because we cannot resolve how the particle spins around its major axis. 
We therefore plan to perform corresponding experiments using micron-sized triangular platelets.  We expect that it will be possible to resolve the tumbling and spinning dynamics of these particles using the methods described in this paper.

\begin{acknowledgments}
We thank S. Gustafsson (Chalmers) for taking the electron-microscope image Fig.~\ref{fig:particle_image}(a) and for granting us permission to use the image in this paper.  B. Mehlig gratefully acknowledges discussions with S. \"Ostlund (University of Gothenburg) on the implications of time-reversal symmetry for the surface-of-section dynamics shown in Fig.~\ref{fig:poincares}.
This work was supported by grants from the Carl Trygger Foundation for Scientific Research  and the Swedish Research Council. Support from the MPNS COST Action MP1305 \lq Flowing matter\rq{} is gratefully acknowledged.
\end{acknowledgments}

\providecommand{\noopsort}[1]{}\providecommand{\singleletter}[1]{#1}%

\begin{thebibliography}{47}%
\makeatletter
\providecommand \@ifxundefined [1]{%
 \@ifx{#1\undefined}
}%
\providecommand \@ifnum [1]{%
 \ifnum #1\expandafter \@firstoftwo
 \else \expandafter \@secondoftwo
 \fi
}%
\providecommand \@ifx [1]{%
 \ifx #1\expandafter \@firstoftwo
 \else \expandafter \@secondoftwo
 \fi
}%
\providecommand \natexlab [1]{#1}%
\providecommand \enquote  [1]{``#1''}%
\providecommand \bibnamefont  [1]{#1}%
\providecommand \bibfnamefont [1]{#1}%
\providecommand \citenamefont [1]{#1}%
\providecommand \href@noop [0]{\@secondoftwo}%
\providecommand \href [0]{\begingroup \@sanitize@url \@href}%
\providecommand \@href[1]{\@@startlink{#1}\@@href}%
\providecommand \@@href[1]{\endgroup#1\@@endlink}%
\providecommand \@sanitize@url [0]{\catcode `\\12\catcode `\$12\catcode
  `\&12\catcode `\#12\catcode `\^12\catcode `\_12\catcode `\%12\relax}%
\providecommand \@@startlink[1]{}%
\providecommand \@@endlink[0]{}%
\providecommand \url  [0]{\begingroup\@sanitize@url \@url }%
\providecommand \@url [1]{\endgroup\@href {#1}{\urlprefix }}%
\providecommand \urlprefix  [0]{URL }%
\providecommand \Eprint [0]{\href }%
\providecommand \doibase [0]{http://dx.doi.org/}%
\providecommand \selectlanguage [0]{\@gobble}%
\providecommand \bibinfo  [0]{\@secondoftwo}%
\providecommand \bibfield  [0]{\@secondoftwo}%
\providecommand \translation [1]{[#1]}%
\providecommand \BibitemOpen [0]{}%
\providecommand \bibitemStop [0]{}%
\providecommand \bibitemNoStop [0]{.\EOS\space}%
\providecommand \EOS [0]{\spacefactor3000\relax}%
\providecommand \BibitemShut  [1]{\csname bibitem#1\endcsname}%
\let\auto@bib@innerbib\@empty
\bibitem [{\citenamefont {Hinch}\ and\ \citenamefont {Leal}(1979)}]{Hinch1979}%
  \BibitemOpen
  \bibfield  {author} {\bibinfo {author} {\bibfnamefont {E.~J.}\ \bibnamefont
  {Hinch}}\ and\ \bibinfo {author} {\bibfnamefont {L.~G.}\ \bibnamefont
  {Leal}},\ }\bibfield  {title} {\enquote {\bibinfo {title} {Rotation of small
  non-axisymmetric particles in a simple shear flow},}\ }\href@noop {}
  {\bibfield  {journal} {\bibinfo  {journal} {J. Fluid Mech.}\ }\textbf
  {\bibinfo {volume} {92}},\ \bibinfo {pages} {591--608} (\bibinfo {year}
  {1979})}\BibitemShut {NoStop}%
\bibitem [{\citenamefont {Yarin}, \citenamefont {Gottlieb},\ and\ \citenamefont
  {Roisman}(1997)}]{Yarin1997}%
  \BibitemOpen
  \bibfield  {author} {\bibinfo {author} {\bibfnamefont {A.~L.}\ \bibnamefont
  {Yarin}}, \bibinfo {author} {\bibfnamefont {O.}~\bibnamefont {Gottlieb}}, \
  and\ \bibinfo {author} {\bibfnamefont {I.~V.}\ \bibnamefont {Roisman}},\
  }\bibfield  {title} {\enquote {\bibinfo {title} {Chaotic rotation of triaxial
  ellipsoids in simple shear flow},}\ }\href@noop {} {\bibfield  {journal}
  {\bibinfo  {journal} {J. Fluid Mech.}\ }\textbf {\bibinfo {volume} {340}},\
  \bibinfo {pages} {83--100} (\bibinfo {year} {1997})}\BibitemShut {NoStop}%
\bibitem [{\citenamefont {Happel}\ and\ \citenamefont
  {Brenner}(1965)}]{Happel1965}%
  \BibitemOpen
  \bibfield  {author} {\bibinfo {author} {\bibfnamefont {J.~R.}\ \bibnamefont
  {Happel}}\ and\ \bibinfo {author} {\bibfnamefont {H.}~\bibnamefont
  {Brenner}},\ }\href@noop {} {\emph {\bibinfo {title} {Low {R}eynolds Number
  Hydrodynamics: With Special Applications to Particulate Media}}}\ (\bibinfo
  {publisher} {Springer},\ \bibinfo {year} {1965})\BibitemShut {NoStop}%
\bibitem [{\citenamefont {Kim}\ and\ \citenamefont {Karrila}(1991)}]{Kim1991}%
  \BibitemOpen
  \bibfield  {author} {\bibinfo {author} {\bibfnamefont {S.}~\bibnamefont
  {Kim}}\ and\ \bibinfo {author} {\bibfnamefont {S.~J.}\ \bibnamefont
  {Karrila}},\ }\href@noop {} {\emph {\bibinfo {title} {Microhydrodynamics:
  principles and selected applications}}},\ Butterworth-Heinemann series in
  {C}hemical {E}ngineering\ (\bibinfo  {publisher} {Butterworth-Heinemann},\
  \bibinfo {address} {Boston},\ \bibinfo {year} {1991})\BibitemShut {NoStop}%
\bibitem [{\citenamefont {Jeffery}(1922)}]{Jeffery1922}%
  \BibitemOpen
  \bibfield  {author} {\bibinfo {author} {\bibfnamefont {G.~B.}\ \bibnamefont
  {Jeffery}},\ }\bibfield  {title} {\enquote {\bibinfo {title} {The {M}otion of
  {E}llipsoidal {P}articles {I}mmersed in a {V}iscous {F}luid},}\ }\href@noop
  {} {\bibfield  {journal} {\bibinfo  {journal} {Proc. R. Soc. A}\ }\textbf
  {\bibinfo {volume} {102}},\ \bibinfo {pages} {161--179} (\bibinfo {year}
  {1922})}\BibitemShut {NoStop}%
\bibitem [{\citenamefont {Bretherton}(1962)}]{Bretherton1962}%
  \BibitemOpen
  \bibfield  {author} {\bibinfo {author} {\bibfnamefont {F.~P.}\ \bibnamefont
  {Bretherton}},\ }\bibfield  {title} {\enquote {\bibinfo {title} {The motion
  of rigid particles in a shear flow at low {R}eynolds number},}\ }\href@noop
  {} {\bibfield  {journal} {\bibinfo  {journal} {J. Fluid Mech.}\ }\textbf
  {\bibinfo {volume} {14}},\ \bibinfo {pages} {284--304} (\bibinfo {year}
  {1962})}\BibitemShut {NoStop}%
\bibitem [{\citenamefont {Taylor}(1923)}]{Taylor1923}%
  \BibitemOpen
  \bibfield  {author} {\bibinfo {author} {\bibfnamefont {G.~I.}\ \bibnamefont
  {Taylor}},\ }\bibfield  {title} {\enquote {\bibinfo {title} {The motion of
  ellipsoidal particles in a viscous fluid},}\ }\href@noop {} {\bibfield
  {journal} {\bibinfo  {journal} {Proc. R. Soc. Lond. A}\ }\textbf {\bibinfo
  {volume} {103}},\ \bibinfo {pages} {58--61} (\bibinfo {year}
  {1923})}\BibitemShut {NoStop}%
\bibitem [{\citenamefont {Karnis}, \citenamefont {Goldsmith},\ and\
  \citenamefont {Mason}(1966)}]{Karnis1966}%
  \BibitemOpen
  \bibfield  {author} {\bibinfo {author} {\bibfnamefont {A.}~\bibnamefont
  {Karnis}}, \bibinfo {author} {\bibfnamefont {H.~L.}\ \bibnamefont
  {Goldsmith}}, \ and\ \bibinfo {author} {\bibfnamefont {S.~G.}\ \bibnamefont
  {Mason}},\ }\bibfield  {title} {\enquote {\bibinfo {title} {The flow of
  suspensions through tubes. V. Inertial effects},}\ }\href@noop {} {\bibfield
  {journal} {\bibinfo  {journal} {J. Chem. Eng.}\ }\textbf {\bibinfo {volume}
  {44}},\ \bibinfo {pages} {181--193} (\bibinfo {year} {1966})}\BibitemShut
  {NoStop}%
\bibitem [{\citenamefont {Subramanian}\ and\ \citenamefont
  {Koch}(2005)}]{subramanian2005}%
  \BibitemOpen
  \bibfield  {author} {\bibinfo {author} {\bibfnamefont {G.}~\bibnamefont
  {Subramanian}}\ and\ \bibinfo {author} {\bibfnamefont {D.~L.}\ \bibnamefont
  {Koch}},\ }\bibfield  {title} {\enquote {\bibinfo {title} {Inertial effects
  on fibre motion in simple shear flow},}\ }\href@noop {} {\bibfield  {journal}
  {\bibinfo  {journal} {J. Fluid Mech.}\ }\textbf {\bibinfo {volume} {535}},\
  \bibinfo {pages} {383--414} (\bibinfo {year} {2005})}\BibitemShut {NoStop}%
\bibitem [{\citenamefont {Einarsson}\ \emph
  {et~al.}(2015{\natexlab{a}})\citenamefont {Einarsson}, \citenamefont
  {Candelier}, \citenamefont {Lundell}, \citenamefont {Angilella},\ and\
  \citenamefont {Mehlig}}]{einarsson2015a}%
  \BibitemOpen
  \bibfield  {author} {\bibinfo {author} {\bibfnamefont {J.}~\bibnamefont
  {Einarsson}}, \bibinfo {author} {\bibfnamefont {F.}~\bibnamefont
  {Candelier}}, \bibinfo {author} {\bibfnamefont {F.}~\bibnamefont {Lundell}},
  \bibinfo {author} {\bibfnamefont {J.}~\bibnamefont {Angilella}}, \ and\
  \bibinfo {author} {\bibfnamefont {B.}~\bibnamefont {Mehlig}},\ }\bibfield
  {title} {\enquote {\bibinfo {title} {Rotation of a spheroid in a simple shear
  at small {R}eynolds number},}\ }\href@noop {} {\bibfield  {journal} {\bibinfo
   {journal} {Phys. Fluids}\ }\textbf {\bibinfo {volume} {27}},\ \bibinfo
  {pages} {063301} (\bibinfo {year} {2015}{\natexlab{a}})}\BibitemShut
  {NoStop}%
\bibitem [{\citenamefont {Einarsson}\ \emph
  {et~al.}(2015{\natexlab{b}})\citenamefont {Einarsson}, \citenamefont
  {Candelier}, \citenamefont {Lundell}, \citenamefont {Angilella},\ and\
  \citenamefont {Mehlig}}]{einarsson2015b}%
  \BibitemOpen
  \bibfield  {author} {\bibinfo {author} {\bibfnamefont {J.}~\bibnamefont
  {Einarsson}}, \bibinfo {author} {\bibfnamefont {F.}~\bibnamefont
  {Candelier}}, \bibinfo {author} {\bibfnamefont {F.}~\bibnamefont {Lundell}},
  \bibinfo {author} {\bibfnamefont {J.}~\bibnamefont {Angilella}}, \ and\
  \bibinfo {author} {\bibfnamefont {B.}~\bibnamefont {Mehlig}},\ }\bibfield
  {title} {\enquote {\bibinfo {title} {Effect of weak fluid inertia upon
  {Jeffery} orbits},}\ }\href@noop {} {\bibfield  {journal} {\bibinfo
  {journal} {Phys. Rev. E}\ }\textbf {\bibinfo {volume} {91}},\ \bibinfo
  {pages} {041002(R)} (\bibinfo {year} {2015}{\natexlab{b}})}\BibitemShut
  {NoStop}%
\bibitem [{\citenamefont {Candelier}\ \emph {et~al.}(2015)\citenamefont
  {Candelier}, \citenamefont {Einarsson}, \citenamefont {Lundell},
  \citenamefont {Mehlig},\ and\ \citenamefont {Angilella}}]{candelier2015}%
  \BibitemOpen
  \bibfield  {author} {\bibinfo {author} {\bibfnamefont {F.}~\bibnamefont
  {Candelier}}, \bibinfo {author} {\bibfnamefont {J.}~\bibnamefont
  {Einarsson}}, \bibinfo {author} {\bibfnamefont {F.}~\bibnamefont {Lundell}},
  \bibinfo {author} {\bibfnamefont {B.}~\bibnamefont {Mehlig}}, \ and\ \bibinfo
  {author} {\bibfnamefont {J.}~\bibnamefont {Angilella}},\ }\bibfield  {title}
  {\enquote {\bibinfo {title} {The role of inertia for the rotation of a nearly
  spherical particle in a general linear flow},}\ }\href@noop {} {\bibfield
  {journal} {\bibinfo  {journal} {Phys. Rev. E}\ }\textbf {\bibinfo {volume}
  {91}},\ \bibinfo {pages} {053023} (\bibinfo {year} {2015})}\BibitemShut
  {NoStop}%
\bibitem [{\citenamefont {Brenner}(1974)}]{Brenner1974}%
  \BibitemOpen
  \bibfield  {author} {\bibinfo {author} {\bibfnamefont {H.}~\bibnamefont
  {Brenner}},\ }\bibfield  {title} {\enquote {\bibinfo {title} {Rheology of a
  dilute suspension of axisymmetric {B}rownian particles},}\ }\href {\doibase
  10.1016/0301-9322(74)90018-4} {\bibfield  {journal} {\bibinfo  {journal}
  {International Journal of Multiphase Flow}\ }\textbf {\bibinfo {volume}
  {1}},\ \bibinfo {pages} {195--341} (\bibinfo {year} {1974})}\BibitemShut
  {NoStop}%
\bibitem [{\citenamefont {Hinch}\ and\ \citenamefont {Leal}(1972)}]{Hinch1972}%
  \BibitemOpen
  \bibfield  {author} {\bibinfo {author} {\bibfnamefont {E.~J.}\ \bibnamefont
  {Hinch}}\ and\ \bibinfo {author} {\bibfnamefont {L.~G.}\ \bibnamefont
  {Leal}},\ }\bibfield  {title} {\enquote {\bibinfo {title} {The effect of
  {B}rownian motion on the rheological properties of a suspension of
  non-spherical particles},}\ }\href {\doibase 10.1017/S002211207200271X}
  {\bibfield  {journal} {\bibinfo  {journal} {Journal of Fluid Mechanics}\
  }\textbf {\bibinfo {volume} {52}},\ \bibinfo {pages} {683--712} (\bibinfo
  {year} {1972})}\BibitemShut {NoStop}%
\bibitem [{\citenamefont {Gierszewski}\ and\ \citenamefont
  {Chaffey}(1978)}]{Gierszewski1978}%
  \BibitemOpen
  \bibfield  {author} {\bibinfo {author} {\bibfnamefont {P.~J.}\ \bibnamefont
  {Gierszewski}}\ and\ \bibinfo {author} {\bibfnamefont {C.~E.}\ \bibnamefont
  {Chaffey}},\ }\bibfield  {title} {\enquote {\bibinfo {title} {Rotation of an
  isolated triaxial ellipsoid suspended in slow viscous flow},}\ }\href@noop {}
  {\bibfield  {journal} {\bibinfo  {journal} {Can.J. Phys.}\ }\textbf {\bibinfo
  {volume} {56}},\ \bibinfo {pages} {6--11} (\bibinfo {year}
  {1978})}\BibitemShut {NoStop}%
\bibitem [{\citenamefont {Petrie}(1999)}]{Petrie1999}%
  \BibitemOpen
  \bibfield  {author} {\bibinfo {author} {\bibfnamefont {C.~J.}\ \bibnamefont
  {Petrie}},\ }\bibfield  {title} {\enquote {\bibinfo {title} {The rheology of
  fibre suspensions},}\ }\href@noop {} {\bibfield  {journal} {\bibinfo
  {journal} {Journal of Non-Newtonian Fluid Mechanics}\ }\textbf {\bibinfo
  {volume} {87}},\ \bibinfo {pages} {369 -- 402} (\bibinfo {year}
  {1999})}\BibitemShut {NoStop}%
\bibitem [{\citenamefont {Parsa}\ \emph {et~al.}(2012)\citenamefont {Parsa},
  \citenamefont {Calzavarini}, \citenamefont {Toschi},\ and\ \citenamefont
  {Voth}}]{Par12}%
  \BibitemOpen
  \bibfield  {author} {\bibinfo {author} {\bibfnamefont {S.}~\bibnamefont
  {Parsa}}, \bibinfo {author} {\bibfnamefont {E.}~\bibnamefont {Calzavarini}},
  \bibinfo {author} {\bibfnamefont {F.}~\bibnamefont {Toschi}}, \ and\ \bibinfo
  {author} {\bibfnamefont {G.~A.}\ \bibnamefont {Voth}},\ }\bibfield  {title}
  {\enquote {\bibinfo {title} {Rotation rate of rods in turbulent fluid
  flow},}\ }\href@noop {} {\bibfield  {journal} {\bibinfo  {journal} {Phys.
  Rev. Lett.}\ }\textbf {\bibinfo {volume} {109}},\ \bibinfo {pages} {134501}
  (\bibinfo {year} {2012})}\BibitemShut {NoStop}%
\bibitem [{\citenamefont {Pumir}\ and\ \citenamefont
  {Wilkinson}(2011)}]{Pum11}%
  \BibitemOpen
  \bibfield  {author} {\bibinfo {author} {\bibfnamefont {A.}~\bibnamefont
  {Pumir}}\ and\ \bibinfo {author} {\bibfnamefont {M.}~\bibnamefont
  {Wilkinson}},\ }\bibfield  {title} {\enquote {\bibinfo {title} {Orientation
  statistics of small particles in turbulence},}\ }\href@noop {} {\bibfield
  {journal} {\bibinfo  {journal} {NJP}\ }\textbf {\bibinfo {volume} {13}},\
  \bibinfo {pages} {093030} (\bibinfo {year} {2011})}\BibitemShut {NoStop}%
\bibitem [{\citenamefont {Gustavsson}, \citenamefont {Einarsson},\ and\
  \citenamefont {Mehlig}(2014)}]{Gus14}%
  \BibitemOpen
  \bibfield  {author} {\bibinfo {author} {\bibfnamefont {K.}~\bibnamefont
  {Gustavsson}}, \bibinfo {author} {\bibfnamefont {J.}~\bibnamefont
  {Einarsson}}, \ and\ \bibinfo {author} {\bibfnamefont {B.}~\bibnamefont
  {Mehlig}},\ }\bibfield  {title} {\enquote {\bibinfo {title} {Tumbling of
  small axisymmetric particles in random and turbulent flows},}\ }\href@noop {}
  {\bibfield  {journal} {\bibinfo  {journal} {Phys. Rev. Lett.}\ }\textbf
  {\bibinfo {volume} {112}},\ \bibinfo {pages} {014501} (\bibinfo {year}
  {2014})}\BibitemShut {NoStop}%
\bibitem [{\citenamefont {Ni}, \citenamefont {Ouelette},\ and\ \citenamefont
  {Voth}(2014)}]{Ni14}%
  \BibitemOpen
  \bibfield  {author} {\bibinfo {author} {\bibfnamefont {R.}~\bibnamefont
  {Ni}}, \bibinfo {author} {\bibfnamefont {N.~T.}\ \bibnamefont {Ouelette}}, \
  and\ \bibinfo {author} {\bibfnamefont {G.~A.}\ \bibnamefont {Voth}},\
  }\bibfield  {title} {\enquote {\bibinfo {title} {Alignment of vorticity and
  rods with {L}agrangian fluid stretching in turbulence},}\ }\href@noop {}
  {\bibfield  {journal} {\bibinfo  {journal} {J. Fluid Mech.}\ }\textbf
  {\bibinfo {volume} {743}},\ \bibinfo {pages} {R3} (\bibinfo {year}
  {2014})}\BibitemShut {NoStop}%
\bibitem [{\citenamefont {Chevillard}\ and\ \citenamefont
  {Meneveau}(2013)}]{Che13}%
  \BibitemOpen
  \bibfield  {author} {\bibinfo {author} {\bibfnamefont {L.}~\bibnamefont
  {Chevillard}}\ and\ \bibinfo {author} {\bibfnamefont {C.}~\bibnamefont
  {Meneveau}},\ }\bibfield  {title} {\enquote {\bibinfo {title} {Orientation
  dynamics of small, triaxial-ellipsoidal particles in isotropic turbulence},}\
  }\href@noop {} {\bibfield  {journal} {\bibinfo  {journal} {J. Fluid Mech.}\
  }\textbf {\bibinfo {volume} {737}},\ \bibinfo {pages} {571} (\bibinfo {year}
  {2013})}\BibitemShut {NoStop}%
\bibitem [{\citenamefont {Byron}\ \emph {et~al.}(2015)\citenamefont {Byron},
  \citenamefont {Einarsson}, \citenamefont {Gustavsson}, \citenamefont {Voth},
  \citenamefont {Mehlig},\ and\ \citenamefont {Variano}}]{Byron2015}%
  \BibitemOpen
  \bibfield  {author} {\bibinfo {author} {\bibfnamefont {M.}~\bibnamefont
  {Byron}}, \bibinfo {author} {\bibfnamefont {J.}~\bibnamefont {Einarsson}},
  \bibinfo {author} {\bibfnamefont {K.}~\bibnamefont {Gustavsson}}, \bibinfo
  {author} {\bibfnamefont {G.}~\bibnamefont {Voth}}, \bibinfo {author}
  {\bibfnamefont {B.}~\bibnamefont {Mehlig}}, \ and\ \bibinfo {author}
  {\bibfnamefont {E.}~\bibnamefont {Variano}},\ }\bibfield  {title} {\enquote
  {\bibinfo {title} {Shape-dependence of particle rotation in isotropic
  turbulence},}\ }\href@noop {} {\bibfield  {journal} {\bibinfo  {journal}
  {Phys. Fluids}\ }\textbf {\bibinfo {volume} {\obs{27}}},\ \bibinfo {pages}
  {\obs{035101}} (\bibinfo {year} {2015})}\BibitemShut {NoStop}%
\bibitem [{\citenamefont {Wilkinson}, \citenamefont {Bezuglyy},\ and\
  \citenamefont {Mehlig}(2009)}]{Wil09}%
  \BibitemOpen
  \bibfield  {author} {\bibinfo {author} {\bibfnamefont {M.}~\bibnamefont
  {Wilkinson}}, \bibinfo {author} {\bibfnamefont {V.}~\bibnamefont {Bezuglyy}},
  \ and\ \bibinfo {author} {\bibfnamefont {B.}~\bibnamefont {Mehlig}},\
  }\bibfield  {title} {\enquote {\bibinfo {title} {Fingerprints of random
  flows},}\ }\href@noop {} {\bibfield  {journal} {\bibinfo  {journal} {Phys.
  Fluids}\ }\textbf {\bibinfo {volume} {21}},\ \bibinfo {pages} {043304}
  (\bibinfo {year} {2009})}\BibitemShut {NoStop}%
\bibitem [{\citenamefont {Bezuglyy}, \citenamefont {Mehlig},\ and\
  \citenamefont {Wilkinson}(2010)}]{Wil10a}%
  \BibitemOpen
  \bibfield  {author} {\bibinfo {author} {\bibfnamefont {V.}~\bibnamefont
  {Bezuglyy}}, \bibinfo {author} {\bibfnamefont {B.}~\bibnamefont {Mehlig}}, \
  and\ \bibinfo {author} {\bibfnamefont {M.}~\bibnamefont {Wilkinson}},\
  }\bibfield  {title} {\enquote {\bibinfo {title} {Poincar\'e indices of
  rheoscopic visualisations},}\ }\href@noop {} {\bibfield  {journal} {\bibinfo
  {journal} {Europhys. Lett.}\ }\textbf {\bibinfo {volume} {89}},\ \bibinfo
  {pages} {34003} (\bibinfo {year} {2010})}\BibitemShut {NoStop}%
\bibitem [{\citenamefont {Wilkinson}, \citenamefont {Bezuglyy},\ and\
  \citenamefont {Mehlig}(2011)}]{Wil11}%
  \BibitemOpen
  \bibfield  {author} {\bibinfo {author} {\bibfnamefont {M.}~\bibnamefont
  {Wilkinson}}, \bibinfo {author} {\bibfnamefont {V.}~\bibnamefont {Bezuglyy}},
  \ and\ \bibinfo {author} {\bibfnamefont {B.}~\bibnamefont {Mehlig}},\
  }\bibfield  {title} {\enquote {\bibinfo {title} {Emergent order in rheoscopic
  swirls},}\ }\href@noop {} {\bibfield  {journal} {\bibinfo  {journal} {J.
  Fluid Mech.}\ }\textbf {\bibinfo {volume} {667}},\ \bibinfo {pages} {158}
  (\bibinfo {year} {2011})}\BibitemShut {NoStop}%
\bibitem [{\citenamefont {Marchioli}, \citenamefont {Fantoni},\ and\
  \citenamefont {Soldati}(2010)}]{Marchioli2010}%
  \BibitemOpen
  \bibfield  {author} {\bibinfo {author} {\bibfnamefont {C.}~\bibnamefont
  {Marchioli}}, \bibinfo {author} {\bibfnamefont {M.}~\bibnamefont {Fantoni}},
  \ and\ \bibinfo {author} {\bibfnamefont {A.}~\bibnamefont {Soldati}},\
  }\bibfield  {title} {\enquote {\bibinfo {title} {Orientation, distribution,
  and deposition of elongated, inertial fibers in turbulent channel flow},}\
  }\href@noop {} {\bibfield  {journal} {\bibinfo  {journal} {Phys. Fluids}\
  }\textbf {\bibinfo {volume} {22}},\ \bibinfo {pages} {033301} (\bibinfo
  {year} {2010})}\BibitemShut {NoStop}%
\bibitem [{\citenamefont {Einarsson}\ \emph {et~al.}(2013)\citenamefont
  {Einarsson}, \citenamefont {Johansson}, \citenamefont {Mahato}, \citenamefont
  {Mishra}, \citenamefont {Angilella}, \citenamefont {Hanstorp},\ and\
  \citenamefont {Mehlig}}]{Einarsson2013a}%
  \BibitemOpen
  \bibfield  {author} {\bibinfo {author} {\bibfnamefont {J.}~\bibnamefont
  {Einarsson}}, \bibinfo {author} {\bibfnamefont {A.}~\bibnamefont
  {Johansson}}, \bibinfo {author} {\bibfnamefont {S.~K.}\ \bibnamefont
  {Mahato}}, \bibinfo {author} {\bibfnamefont {Y.~N.}\ \bibnamefont {Mishra}},
  \bibinfo {author} {\bibfnamefont {J.}~\bibnamefont {Angilella}}, \bibinfo
  {author} {\bibfnamefont {D.}~\bibnamefont {Hanstorp}}, \ and\ \bibinfo
  {author} {\bibfnamefont {B.}~\bibnamefont {Mehlig}},\ }\bibfield  {title}
  {\enquote {\bibinfo {title} {Periodic and aperiodic tumbling of microrods
  advected in a microchannel flow},}\ }\href@noop {} {\bibfield  {journal}
  {\bibinfo  {journal} {Acta Mechanica}\ }\textbf {\bibinfo {volume} {224}},\
  \bibinfo {pages} {2281--2289} (\bibinfo {year} {2013})}\BibitemShut {NoStop}%
\bibitem [{\citenamefont {Challabotla}, \citenamefont {Zhao},\ and\
  \citenamefont {Andersson}(2015)}]{Challabotla2015}%
  \BibitemOpen
  \bibfield  {author} {\bibinfo {author} {\bibfnamefont {N.~R.}\ \bibnamefont
  {Challabotla}}, \bibinfo {author} {\bibfnamefont {L.}~\bibnamefont {Zhao}}, \
  and\ \bibinfo {author} {\bibfnamefont {H.}~\bibnamefont {Andersson}},\
  }\bibfield  {title} {\enquote {\bibinfo {title} {Orientation and rotation of
  inertial disk particles in wall turbulence},}\ }\href@noop {} {\bibfield
  {journal} {\bibinfo  {journal} {J. Fluid Mech.}\ }\textbf {\bibinfo {volume}
  {766}},\ \bibinfo {pages} {R2} (\bibinfo {year} {2015})}\BibitemShut
  {NoStop}%
\bibitem [{\citenamefont {Eirich}, \citenamefont {Margaretha},\ and\
  \citenamefont {Bunzl}(1936)}]{Eirich1936}%
  \BibitemOpen
  \bibfield  {author} {\bibinfo {author} {\bibfnamefont {F.}~\bibnamefont
  {Eirich}}, \bibinfo {author} {\bibfnamefont {H.}~\bibnamefont {Margaretha}},
  \ and\ \bibinfo {author} {\bibfnamefont {M.}~\bibnamefont {Bunzl}},\
  }\bibfield  {title} {\enquote {\bibinfo {title} {Untersuchungen {\"{u}}ber
  die {V}iskosit{\"{a}}t von {S}uspensionen und {L\"{o}}sungen},}\ }\href@noop
  {} {\bibfield  {journal} {\bibinfo  {journal} {Kolloid-Zeitschrift}\ }\textbf
  {\bibinfo {volume} {75}},\ \bibinfo {pages} {20--37} (\bibinfo {year}
  {1936})}\BibitemShut {NoStop}%
\bibitem [{\citenamefont {Binder}(1939)}]{Binder1939}%
  \BibitemOpen
  \bibfield  {author} {\bibinfo {author} {\bibfnamefont {R.~C.}\ \bibnamefont
  {Binder}},\ }\bibfield  {title} {\enquote {\bibinfo {title} {The motion of
  cylindrical particles in viscous flow},}\ }\href {\doibase 10.1063/1.1707254}
  {\bibfield  {journal} {\bibinfo  {journal} {Journal of Applied Physics}\
  }\textbf {\bibinfo {volume} {10}},\ \bibinfo {pages} {711--713} (\bibinfo
  {year} {1939})}\BibitemShut {NoStop}%
\bibitem [{\citenamefont {Trevelyan}\ and\ \citenamefont
  {Mason}(1951)}]{Trevelyan1951}%
  \BibitemOpen
  \bibfield  {author} {\bibinfo {author} {\bibfnamefont {J.}~\bibnamefont
  {Trevelyan}}\ and\ \bibinfo {author} {\bibfnamefont {S.~G.}\ \bibnamefont
  {Mason}},\ }\bibfield  {title} {\enquote {\bibinfo {title} {Particle motion
  in a sheared suspensions. {I.} {R}otations},}\ }\href@noop {} {\bibfield
  {journal} {\bibinfo  {journal} {J. Colloi. Sci}\ }\textbf {\bibinfo {volume}
  {6}},\ \bibinfo {pages} {354--367} (\bibinfo {year} {1951})}\BibitemShut
  {NoStop}%
\bibitem [{\citenamefont {Mason}\ and\ \citenamefont
  {Manley}(1956)}]{Mason1956}%
  \BibitemOpen
  \bibfield  {author} {\bibinfo {author} {\bibfnamefont {S.~G.}\ \bibnamefont
  {Mason}}\ and\ \bibinfo {author} {\bibfnamefont {R.~S.~J.}\ \bibnamefont
  {Manley}},\ }\bibfield  {title} {\enquote {\bibinfo {title} {Particle motions
  in sheared suspensions: Orientations and interactions of rigid rods},}\
  }\href@noop {} {\bibfield  {journal} {\bibinfo  {journal} {Proceedings of the
  Royal Society of London. Series A, Mathematical and Physical Sciences}\
  }\textbf {\bibinfo {volume} {238}},\ \bibinfo {pages} {117--131} (\bibinfo
  {year} {1956})}\BibitemShut {NoStop}%
\bibitem [{\citenamefont {Bartok}\ and\ \citenamefont
  {Mason}(1957)}]{Bartok1957}%
  \BibitemOpen
  \bibfield  {author} {\bibinfo {author} {\bibfnamefont {W.}~\bibnamefont
  {Bartok}}\ and\ \bibinfo {author} {\bibfnamefont {S.~G.}\ \bibnamefont
  {Mason}},\ }\bibfield  {title} {\enquote {\bibinfo {title} {Particle motion
  in sheared suspensions},}\ }\href@noop {} {\bibfield  {journal} {\bibinfo
  {journal} {J. Colloid Sci.}\ }\textbf {\bibinfo {volume} {12}},\ \bibinfo
  {pages} {243--262} (\bibinfo {year} {1957})}\BibitemShut {NoStop}%
\bibitem [{\citenamefont {Goldsmith}\ and\ \citenamefont
  {Mason}(1961)}]{Goldsmith1961}%
  \BibitemOpen
  \bibfield  {author} {\bibinfo {author} {\bibfnamefont {H.~L.}\ \bibnamefont
  {Goldsmith}}\ and\ \bibinfo {author} {\bibfnamefont {S.~G.}\ \bibnamefont
  {Mason}},\ }\bibfield  {title} {\enquote {\bibinfo {title} {Particle motions
  in sheared suspensions. {XIII}. {T}he spin and rotation of disks},}\
  }\href@noop {} {\bibfield  {journal} {\bibinfo  {journal} {J. Fluid Mech.}\
  }\textbf {\bibinfo {volume} {12}},\ \bibinfo {pages} {88 -- 96} (\bibinfo
  {year} {1962})}\BibitemShut {NoStop}%
\bibitem [{\citenamefont {Goldsmith}\ and\ \citenamefont
  {Mason}(1962)}]{Goldsmith1962}%
  \BibitemOpen
  \bibfield  {author} {\bibinfo {author} {\bibfnamefont {H.~L.}\ \bibnamefont
  {Goldsmith}}\ and\ \bibinfo {author} {\bibfnamefont {S.~G.}\ \bibnamefont
  {Mason}},\ }\bibfield  {title} {\enquote {\bibinfo {title} {The flow of
  suspensions through tubes. {I}. {S}ingle spheres, rods, and discs},}\
  }\href@noop {} {\bibfield  {journal} {\bibinfo  {journal} {J. Colloi. Sci.}\
  }\textbf {\bibinfo {volume} {17}},\ \bibinfo {pages} {448 -- 476} (\bibinfo
  {year} {1962})}\BibitemShut {NoStop}%
\bibitem [{\citenamefont {Anczurowski}\ and\ \citenamefont
  {Mason}(1968)}]{Anczurowski1968}%
  \BibitemOpen
  \bibfield  {author} {\bibinfo {author} {\bibfnamefont {E.}~\bibnamefont
  {Anczurowski}}\ and\ \bibinfo {author} {\bibfnamefont {S.~G.}\ \bibnamefont
  {Mason}},\ }\bibfield  {title} {\enquote {\bibinfo {title} {Particle motions
  in sheared suspensions. {XXIV.} {R}otation of rigid spheriods and
  cylinders},}\ }\href@noop {} {\bibfield  {journal} {\bibinfo  {journal}
  {Transaction of The Society of Rheology}\ }\textbf {\bibinfo {volume} {12}},\
  \bibinfo {pages} {209--215} (\bibinfo {year} {1968})}\BibitemShut {NoStop}%
\bibitem [{\citenamefont {Harris}, \citenamefont {Nawaz},\ and\ \citenamefont
  {Pittman}(1979)}]{Harris1979}%
  \BibitemOpen
  \bibfield  {author} {\bibinfo {author} {\bibfnamefont {J.~B.}\ \bibnamefont
  {Harris}}, \bibinfo {author} {\bibfnamefont {M.}~\bibnamefont {Nawaz}}, \
  and\ \bibinfo {author} {\bibfnamefont {J.~F.~T.}\ \bibnamefont {Pittman}},\
  }\bibfield  {title} {\enquote {\bibinfo {title} {Low-{R}eynolds-number motion
  of particles with two or three perpendicular planes of symmetry},}\ }\href
  {\doibase 10.1017/S0022112079001531} {\bibfield  {journal} {\bibinfo
  {journal} {Journal of Fluid Mechanics}\ }\textbf {\bibinfo {volume} {95}},\
  \bibinfo {pages} {415--429} (\bibinfo {year} {1979})}\BibitemShut {NoStop}%
\bibitem [{\citenamefont {Stover}\ and\ \citenamefont
  {Cohen}(1990)}]{Stover1990}%
  \BibitemOpen
  \bibfield  {author} {\bibinfo {author} {\bibfnamefont {C.~A.}\ \bibnamefont
  {Stover}}\ and\ \bibinfo {author} {\bibfnamefont {C.}~\bibnamefont {Cohen}},\
  }\bibfield  {title} {\enquote {\bibinfo {title} {The motion of rodlike
  particles in the pressure-driven flow between two flat plates},}\ }\href@noop
  {} {\bibfield  {journal} {\bibinfo  {journal} {Rheol Acta}\ }\textbf
  {\bibinfo {volume} {29}},\ \bibinfo {pages} {192--203} (\bibinfo {year}
  {1990})}\BibitemShut {NoStop}%
\bibitem [{\citenamefont {Kaya}\ and\ \citenamefont {Koser}(2009)}]{Kaya2009}%
  \BibitemOpen
  \bibfield  {author} {\bibinfo {author} {\bibfnamefont {T.}~\bibnamefont
  {Kaya}}\ and\ \bibinfo {author} {\bibfnamefont {H.}~\bibnamefont {Koser}},\
  }\bibfield  {title} {\enquote {\bibinfo {title} {Characterization of
  hydrodynamics surface interactions of escherichia coli cell bodies in shear
  flow},}\ }\href@noop {} {\bibfield  {journal} {\bibinfo  {journal} {Phys.
  Rev. Lett.}\ }\textbf {\bibinfo {volume} {103}},\ \bibinfo {pages} {138103}
  (\bibinfo {year} {2009})}\BibitemShut {NoStop}%
\bibitem [{\citenamefont {Alargova}\ \emph {et~al.}(2004)\citenamefont
  {Alargova}, \citenamefont {Bhatt}, \citenamefont {Paunov},\ and\
  \citenamefont {Velev}}]{alargova2004}%
  \BibitemOpen
  \bibfield  {author} {\bibinfo {author} {\bibfnamefont {R.}~\bibnamefont
  {Alargova}}, \bibinfo {author} {\bibfnamefont {K.}~\bibnamefont {Bhatt}},
  \bibinfo {author} {\bibfnamefont {V.}~\bibnamefont {Paunov}}, \ and\ \bibinfo
  {author} {\bibfnamefont {O.}~\bibnamefont {Velev}},\ }\bibfield  {title}
  {\enquote {\bibinfo {title} {Scalable synthesis of a new class of polymer
  microrods by a liquid-liquid dispersion technique},}\ }\href@noop {}
  {\bibfield  {journal} {\bibinfo  {journal} {Advanced Materials}\ }\textbf
  {\bibinfo {volume} {16}},\ \bibinfo {pages} {1653--1657} (\bibinfo {year}
  {2004})}\BibitemShut {NoStop}%
\bibitem [{\citenamefont {Lewandowski}\ \emph {et~al.}(2008)\citenamefont
  {Lewandowski}, \citenamefont {Bernate}, \citenamefont {Searson},\ and\
  \citenamefont {Stebe}}]{Lewandowski2008}%
  \BibitemOpen
  \bibfield  {author} {\bibinfo {author} {\bibfnamefont {E.~P.}\ \bibnamefont
  {Lewandowski}}, \bibinfo {author} {\bibfnamefont {J.~A.}\ \bibnamefont
  {Bernate}}, \bibinfo {author} {\bibfnamefont {P.~C.}\ \bibnamefont
  {Searson}}, \ and\ \bibinfo {author} {\bibfnamefont {K.~J.}\ \bibnamefont
  {Stebe}},\ }\bibfield  {title} {\enquote {\bibinfo {title} {Rotation and
  alignment of anisotopic particles on nonplanar interfaces},}\ }\href@noop {}
  {\bibfield  {journal} {\bibinfo  {journal} {Langmuir}\ }\textbf {\bibinfo
  {volume} {24}},\ \bibinfo {pages} {9302--9307} (\bibinfo {year}
  {2008})}\BibitemShut {NoStop}%
\bibitem [{sup()}]{supp}%
  \BibitemOpen
  \href@noop {} {}\bibinfo {note} {Additional experimental data (URL TO BE
  INSERTED BY AIPP)}\BibitemShut {NoStop}%
\bibitem [{\citenamefont {Einarsson}(2013)}]{Einarsson2013b}%
  \BibitemOpen
  \bibfield  {author} {\bibinfo {author} {\bibfnamefont {J.}~\bibnamefont
  {Einarsson}},\ }\href@noop {} {\enquote {\bibinfo {title} {Orientational
  dynamics of small non-spherical particles in fluid flows},}\ } (\bibinfo
  {year} {2013}),\ \bibinfo {note} {{L}icentiate thesis, {D}epartment of
  {P}hysics, {U}niversity of {G}othenburg}\BibitemShut {NoStop}%
\bibitem [{\citenamefont {Strogatz}(1994)}]{Strogatz}%
  \BibitemOpen
  \bibfield  {author} {\bibinfo {author} {\bibfnamefont {S.~H.}\ \bibnamefont
  {Strogatz}},\ }\href@noop {} {\emph {\bibinfo {title} {Nonlinear dynamics and
  {C}haos}}}\ (\bibinfo  {publisher} {Westview Press},\ \bibinfo {year}
  {1994})\BibitemShut {NoStop}%
\bibitem [{\citenamefont {Goldstein}(1980)}]{Goldstein}%
  \BibitemOpen
  \bibfield  {author} {\bibinfo {author} {\bibfnamefont {H.}~\bibnamefont
  {Goldstein}},\ }\href@noop {} {\emph {\bibinfo {title} {Classical
  Mechanics}}}\ (\bibinfo  {publisher} {Addison-Wesley},\ \bibinfo {address}
  {Reading, Massachusetts},\ \bibinfo {year} {1980})\BibitemShut {NoStop}%
\bibitem [{\citenamefont {Politi}, \citenamefont {Oppo},\ and\ \citenamefont
  {Badii}(1986)}]{Politi1986}%
  \BibitemOpen
  \bibfield  {author} {\bibinfo {author} {\bibfnamefont {A.}~\bibnamefont
  {Politi}}, \bibinfo {author} {\bibfnamefont {G.~L.}\ \bibnamefont {Oppo}}, \
  and\ \bibinfo {author} {\bibfnamefont {R.}~\bibnamefont {Badii}},\ }\bibfield
   {title} {\enquote {\bibinfo {title} {Coexistence of conservative and
  dissipative behaviour in reversible dynamical systems},}\ }\href@noop {}
  {\bibfield  {journal} {\bibinfo  {journal} {Phys. Rev. A}\ }\textbf {\bibinfo
  {volume} {33}},\ \bibinfo {pages} {4055} (\bibinfo {year}
  {1986})}\BibitemShut {NoStop}%
\bibitem [{\citenamefont {Lichtenberg}\ and\ \citenamefont
  {Lieberman}(1983)}]{Lichtenberg}%
  \BibitemOpen
  \bibfield  {author} {\bibinfo {author} {\bibfnamefont {A.~J.}\ \bibnamefont
  {Lichtenberg}}\ and\ \bibinfo {author} {\bibfnamefont {M.~A.}\ \bibnamefont
  {Lieberman}},\ }\href@noop {} {\emph {\bibinfo {title} {Regular and
  stochastic motion}}}\ (\bibinfo  {publisher} {Springer},\ \bibinfo {address}
  {New York},\ \bibinfo {year} {1983})\BibitemShut {NoStop}%
\end{thebibliography}
\end{document}